\documentclass{article}
\usepackage{bm,latexsym,amsmath,amssymb,amsfonts,fancyhdr,color,graphicx,multirow,axodraw}
\usepackage[a4paper,bottom=2cm,top=2.5cm,head=5mm,width=18cm,dvipdfm]{geometry}
\usepackage[usenames,dvipsnames,svgnames,table]{xcolor}
\usepackage[ps2pdf,
            colorlinks=true,
            linkcolor=red,
            urlcolor=blue,
            citecolor=green,          
						bookmarks=true,
						bookmarksnumbered=true,
						breaklinks=true,
						pdfpagemode=Fullscreen,
						pdfstartview=FitBH]{hyperref}
\usepackage[dotinlabels]{titletoc}
\usepackage{titlesec}
\usepackage{authblk}

\numberwithin{equation}{section}
\allowdisplaybreaks[4]

\titlelabel{\thetitle.\quad \hspace{-0.8em}}
\titlecontents{section}
              [1.5em]
              {\vspace{4mm} \large \bf}
              {\contentslabel{1em}}
              {\hspace*{-1em}}
              {\titlerule*[.5pc]{.}\contentspage}
\titlecontents{subsection}
              [3.5em]
              {\vspace{2mm}}
              {\contentslabel{1.8em}}
              {\hspace*{.3em}}
              {\titlerule*[.5pc]{.}\contentspage}
\titlecontents{subsubsection}
              [5.5em]
              {\vspace{2mm}}
              {\contentslabel{2.5em}}
              {\hspace*{.3em}}
              {\titlerule*[.5pc]{.}\contentspage}

\hypersetup{ pdfauthor = {Shao-Feng Ge},
	     pdftitle = {Physics Reach of Atmospheric Neutrino Measurements at PINGU}, 
	     pdfsubject = {}, 
             pdfkeywords = {}, 
	     pdfcreator = {LaTeX with hyperref package},
	     pdfproducer = {dvips + ps2pdf} }

\definecolor{gesfpurple}{rgb}{0.47,0.19,0.42}

\definecolor{gesflanse}{rgb}{0.00,0.50,0.50}

\definecolor{gesfblue}{rgb}{0.08,0.42,0.76}
\newcommand{\gblue}[1]{{\color{gesfblue} #1}}
\definecolor{gesfred}{rgb}{1,0,0}
\newcommand{\gred}[1]{{\color{gesfred} #1}}
\definecolor{gesfwhite}{rgb}{1,1,1}

\definecolor{gesfblack}{rgb}{0,0,0}

\newcommand{\gsec}[1]{{\hypersetup{linkcolor=blue}Sec.~\ref{#1}\hypersetup{linkcolor=red}}}
\newcommand{\gfig}[1]{{\hypersetup{linkcolor=violet}Fig.~\ref{#1}\hypersetup{linkcolor=red}}}
\newcommand{\gtab}[1]{{\hypersetup{linkcolor=gesflanse}Table~\ref{#1}\hypersetup{linkcolor=red}}}

\begin{document}

\title{\begin{flushright}
       \mbox{\normalsize KEK-TH-1691}
       \end{flushright}
			 \vskip 20pt
       \textbf{\huge Physics Reach of Atmospheric Neutrino Measurements at PINGU}} 
\author[1]{{\large Shao-Feng Ge} \footnote{gesf02@gmail.com}}
\author[2]{{\large Kaoru Hagiwara} \footnote{kaoru.hagiwara@kek.jp}}
\affil[1]{\small KEK Theory Center, Tsukuba, 305-0801, Japan}
\affil[2]{KEK Theory Center and Sokendai, Tsukuba, 305-0801, Japan}
\date{\today}

\maketitle

\begin{abstract}
The sensitivity of a huge underground water/ice Cherenkov detector, 
such as PINGU in IceCube, to the neutrino mass hierarchy, 
the atmospheric mixing angle and its octant, is studied in detail. 
Based on the event rate decomposition in the propagation basis, 
we illustrate the smearing effects from the neutrino scattering,
the visible energy and zenith angle reconstruction procedures, 
the energy and angular resolutions, and the muon mis-identification rate, 
as well as the impacts of 
systematic errors in the detector resolutions, the muon mis-identification 
rate, and the overall normalization. The sensitivity, especially the 
mass hierarchy sensitivity, can be
enhanced by splitting the muon-like events into two channels, according to 
the event inelasticity. We also show 
that including the cascade events can improve and stabilize the 
sensitivity of the measurements. 

\end{abstract}

\section{Introduction}

The reactor mixing angle, $\theta_{13}$, has been measured in the last
two years. Following T2K \cite{t2k}, MINOS \cite{minos} and Double Chooz 
\cite{doublechooz} indicated that $\theta_{13}$ is nonzero, reaching
$3.5 \, \sigma$ in a combined analysis \cite{Machado:2011ar}.
The conclusive results of a large reactor mixing angle come from Daya Bay 
\cite{dayabay} and RENO \cite{reno} with significance up to
$7.7 \, \sigma$ \cite{dayabay2}. These progresses open up the opportunity 
\cite{Minakata12,Roy12,Pascoli13} to measure the three remaining unknown 
parameters in neutrino oscillation, namely the neutrino mass hierarchy, 
the octant of the atmospheric mixing angle, and the CP phase. 

The neutrino mass hierarchy can be measured \cite{mass1} by 
medium-baseline 
reactor experiments, such as JUNO \cite{juno} and RENO-50 \cite{reno50}, 
atmospheric neutrino experiments, such as PINGU (Precision Icecube Next 
Generation Upgrade) \cite{pingu1}, Hyper-K \cite{sk1,sk2}, INO \cite{ino,ical1}, 
or a liquid argon detector \cite{liquid0,liquid1,liquid2,liquid3}, and accelerator based long-baseline (LBL)
experiments such as NO$\nu$A \cite{nova1,nova2} or 
LBNE \cite{lbne1}. The reactor based experiments focus on the neutrino mass hierarchy 
while atmospheric neutrino experiments can measure both the neutrino mass hierarchy
and the atmospheric mixing angle. In addition to the neutrino mass hierarchy, accelerator 
experiments \cite{accelerator} can measure the CP phase \cite{cp1}, including 
NO$\nu$A \cite{nova1,nova2}, T2K \cite{jhf}, T2HK \cite{sk1,sk2,t2hk}, and LBNE 
\cite{lbne1}.
By splitting the running time between neutrino and antineutrino, it is also 
possible for accelerator experiments to achieve stable sensitivity to the octant 
of the atmospheric mixing angle \cite{split1,cp1,z2}.

For atmospheric neutrino experiments, different ways have been developed to analyse the
oscillation pattern, including oscillogram 
\cite{oscillogram0a,length4,oscillogram0c,oscillogram0d,oscillogram0e,oscillogram0,oscillogram0,oscillogram1} and event-rate decomposition \cite{Ge2013}. 
The former emphasizes the overall structure, especially for the resonance behavior,
while the later is designed for separating the contributions of the three
unknown parameters hidden in the overall pattern and hence is very powerful to 
unveil the trend in $\chi^2$ minimization. In principle, the oscillogram
can also be applied to the coefficients of analytically decomposed terms, 
making their overall pattern explicit. Here, we apply the algorithm 
developed in \cite{Ge2013} on the physics reach of the PINGU experiment.
The sensitivity of PINGU has been explored extensively in the literature
\cite{Smirnov12,Franco13,Ribordy13,Winter13,Blennow1306,icrc555,pingu1306,Ge2013} 
for the neutrino mass hierarchy, and in \cite{Smirnov12,octant1,octant2,Ge2013} for 
the octant of the atmospheric mixing angle. There are some brief discussions on 
the CP phase \cite{Smirnov12,Ohlsson13,Ge2013}.

In this paper, we study in detail the physics reach of atmospheric neutrino
oscillation measurements at the PINGU detector, concerning the sensitivity to 
the neutrino mass hierarchy, the precision on the atmospheric mixing angle and 
the sensitivity to its octant. This paper is organized as follows. In
\gsec{sec:decomposition}, the formalism of event-rate decomposition in 
the propagation basis is summarized. In \gsec{sec:scattering} we present the 
basic features of neutrino scattering, especially the inelasticity distribution 
which can be used to enhance the sensitivity to the neutrino mass hierarchy, 
and reconstruction procedures of the visible energy and zenith angle, 
leaving discussions on the detector resolutions to \gsec{sec:resolution}. 
The impacts of all these smearing effects are studied step by step in 
\gsec{sec:chi2}. Finally in \gsec{sec:sys}, we examine possible impacts of a 
few systematic errors, including those
in energy and angular resolutions, muon mis-identification rate, and
overall normalization, and the conclusion can be found in 
\gsec{sec:conclusion}.

\section{Event-Rate Decomposition in the Propagation Basis}
\label{sec:decomposition}

With the reactor mixing angle being measured, there are three remaining unknowns
in the three-neutrino oscillation, 
namely the neutrino mass hierarchy, the octant of the atmospheric 
mixing angle, $\theta_{\rm a} \equiv \theta_{23}$, and the CP phase, $\delta$. 
Their contributions to neutrino oscillation 
can be analytically decomposed in the propagation basis 
\cite{nupro0,nupro1,nupro2}. This formalism can apply generally and is extremely 
useful for the study of atmospheric neutrino oscillation where the matter 
potential has a complicated structure \cite{prem}.

For completeness, we review the key results of the event-rate decomposition 
\cite{Ge2013} in the propagation basis. By noting that the atmospheric mixing 
angle $\theta_{\rm a}$ and the CP phase $\delta$ dependences of the oscillation
amplitudes do not suffer from the matter effects in the propagation basis, 
one can express analytically their dependences of all the oscillation 
probabilities, and hence the expected event rates in any neutrino 
experiments/observations. The following decompositions for the muon-like 
($\alpha = \mu$) and the cascade ($\alpha = e$) event
rates \footnote{In contrast to the word ``{\it electron-like}'' used in our 
earlier publication \cite{Ge2013}, we use ``{\it cascade}'' instead, following
the IceCube collaboration.} were proposed for phenomenological studies,
\begin{equation}
  \frac {\partial^2 N_\alpha(E_\nu, \cos \theta^\nu_{\rm z})}
				{\partial E_\nu \partial \cos \theta^\nu_{\rm z}} 
=
  N^{(0)}_\alpha 
+ N^{(1)}_\alpha x_{\rm a}
+ N^{(2)}_\alpha \cos \delta'
+ N^{(3)}_\alpha \sin \delta'
+ N^{(4)}_\alpha x_{\rm a} \cos \delta'
+ N^{(5)}_\alpha x^2_{\rm a}
+ N^{(6)}_\alpha \cos^2 \delta'
+ \mathcal O(x^4_{\rm a}) \,,
\label{eq:dNdE}
\end{equation}
where,
\begin{equation}
  x_{\rm a}
\equiv
  \cos 2 \theta_{\rm a}
=
  \cos^2 \theta_{\rm a}
- \sin^2 \theta_{\rm a} \,,
\end{equation}
parametrizes the dependence of the oscillation probabilities on the 
atmospheric mixing angle $\theta_{\rm a}$, taking a positive value for
the lower octant (LO), $\sin^2 \theta_{\rm a} < 0.5$, and a negative value
for the higher octant (HO), $\sin^2 \theta_{\rm a} > 0.5$. The CP phase
$\delta$ dependence appear only through the combinations,
\begin{equation}
  \cos \delta'
\equiv
  \sin 2 \theta_{\rm a} \cos \delta 
=
  \sqrt{1 - x^2_{\rm a}} \cos \delta \,, 
\qquad
  \sin \delta'
\equiv
  \sin 2 \theta_{\rm a} \sin \delta 
=
  \sqrt{1 - x^2_{\rm a}} \sin \delta \,,
\end{equation}
since $\cos \delta$ or $\sin \delta$ are always modulated by a common prefactor
$\sin 2 \theta_{\rm a}$. The accuracy of measuring the CP phase decreases
when the atmospheric mixing angle deviates from its maximal value. Note that
the expansion is up to order $x^2_{\rm a}$ as
the terms of order $x^4_{\rm a}$ are found to be negligibly small \cite{Ge2013}
in the $3 \, \sigma$ allowed range, 
\begin{equation}
  x^2_{\rm a} = 1 - \sin^2 2 \theta_{\rm a} < 0.052 \,.
\end{equation}

For atmospheric neutrino oscillation, 
the coefficients $N^{(k)}_\mu$ and $N^{(k)}_{\rm e}$ ($k = 0, \cdots, 6$) give 
the neutrino energy, $E_\nu$, and the zenith angle, $\cos \theta^\nu_{\rm z}$, 
dependences of the muon-like and cascade event rates, respectively, which
dependent on the solar mixing angle, $\theta_{\rm s} \equiv \theta_{12}$,
the reactor mixing angle, $\theta_{\rm r} \equiv \theta_{13}$, and the two
mass squared differences, $\delta m^2_{\rm a} \equiv m^2_3 - m^2_1$ and
$\delta m^2_{\rm s} \equiv m^2_2 - m^2_1$, and hence on the neutrino mass
hierarchy, normal ($\delta m^2_{\rm a} > 0$) or inverted 
($\delta m^2_{\rm a} < 0$). 
These coefficients are obtained by the convolution integrals,
\begin{eqnarray}
  N^{(k)}_\alpha (E_\nu, \cos \theta^\nu_{\rm z})
& = &
  \sum_{\beta = {\rm e}, \mu}
\left[
  \phi_{\nu_\beta}(E_\nu, \cos \theta^\nu_{\rm z}) \times
  P^{(k)}_{\beta \alpha}(E_\nu, \cos \theta^\nu_{\rm z}) \times
  \sigma_{\nu_\alpha}(E_\nu)
\right.
\nonumber
\\
&&
\hspace{7mm}
\left.
+
  \phi_{\bar \nu_\beta}(E_\nu, \cos \theta^\nu_{\rm z}) \times
  \overline P^{(k)}_{\beta \alpha}(E_\nu, \cos \theta^\nu_{\rm z}) \times
  \sigma_{\bar \nu_\alpha}(E_\nu)
\right]
  \times \rho V_{\rm eff}(E_\nu) \,,
\label{eq:dNdE-integ}
\end{eqnarray} 
where $\phi_{\nu_\beta}$ and $\phi_{\bar \nu_\beta}$ are the $\nu_\beta$
and $\bar \nu_\beta$ fluxes at the South Pole \cite{honda12},
$P^{(k)}_{\beta \alpha}$ and $\overline P^{(k)}_{\beta \alpha}$ are,
respectively, the relevant coefficients of the $\nu_\beta \rightarrow \nu_\alpha$
and $\bar \nu_\beta \rightarrow \bar \nu_\alpha$ oscillation probabilities 
\cite{Ge2013}, $\sigma_{\nu_\alpha}$ and $\sigma_{\bar \nu_\alpha}$ denote
the $\nu_\alpha$ and $\bar \nu_\alpha$ cross sections via the charged current
(CC) scattering off the water target obtained with NEUGEN \cite{neugen}, and 
$\rho V_{\rm eff}$ is the effective fiducial volume of the detector. 
Although the atmospheric neutrino flux for $\nu_\tau$ or 
$\bar \nu_\tau$ is negligibly small \cite{honda12}, we can account for the 
$\nu_\tau$ and $\bar \nu_\tau$ scattering contributions, by using the same
formula (\ref{eq:dNdE-integ}) for $\alpha = \tau$, even though the effective 
fiducial volume for $\tau$-CC and neutral current (NC) events may be much smaller
than those of $\mu$-CC and e-CC events \cite{Tang11}.

The coefficients of those terms independent of CP phase $\delta$, 
namely $N^{(0)}_\alpha$,
$N^{(1)}_\alpha$, and $N^{(5)}_\mu$, are at least one order of magnitude 
larger than those of $\delta$-dependent terms: $N^{(2)}_\alpha$, $N^{(3)}_\alpha$,
and $N^{(4)}_\mu$ are of the order 
$\mathcal O(\delta m^2_{\rm s} / \delta m^2_{\rm a})$, while $N^{(6)}_\mu$
is suppressed by $(\delta m^2_{\rm s} / \delta m^2_{\rm a})^2$ as compared to
the overall rates of $N^{(0)}_\alpha$. Note that the cascade event rates
have no nonlinear dependence on $x_{\rm a}$ and $\cos \delta'$, 
$N^{(4)}_{\rm e} = N^{(5)}_{\rm e} = N^{(6)}_{\rm e} = 0$. The smallness or 
absense of the coefficients $N^{(k)}_\alpha$ for $k = 2, 3, 4, 6$ in 
(\ref{eq:dNdE}) are consequences of the smallness or absense of the 
corresponding oscillation probabilities $P^{(k)}_{\beta \alpha}$ and 
$\overline P^{(k)}_{\beta \alpha}$ for $k = 2, 3, 4, 6$ \cite{Ge2013}.
These features tell that the neutrino mass hierarchy and the atmospheric 
mixing angle can have sizable effects and hence be measured at PINGU, 
but it is more challenging to measure the CP phase. In this paper, we 
concentrate on the sensitivity of PINGU to the neutrino mass hierarchy, its 
expected accuracy of measuring the atmospheric mixing angle $\theta_{\rm a}$,
parametrized as $x_{\rm a}$, and sensitivity to its octant 
($x_{\rm a} < 0 \mbox{ or } x_{\rm a} > 0$).

\begin{figure}[h!]
\centering
\vspace{2mm}
\includegraphics[height=0.32\textwidth,width=7cm,angle=-90]{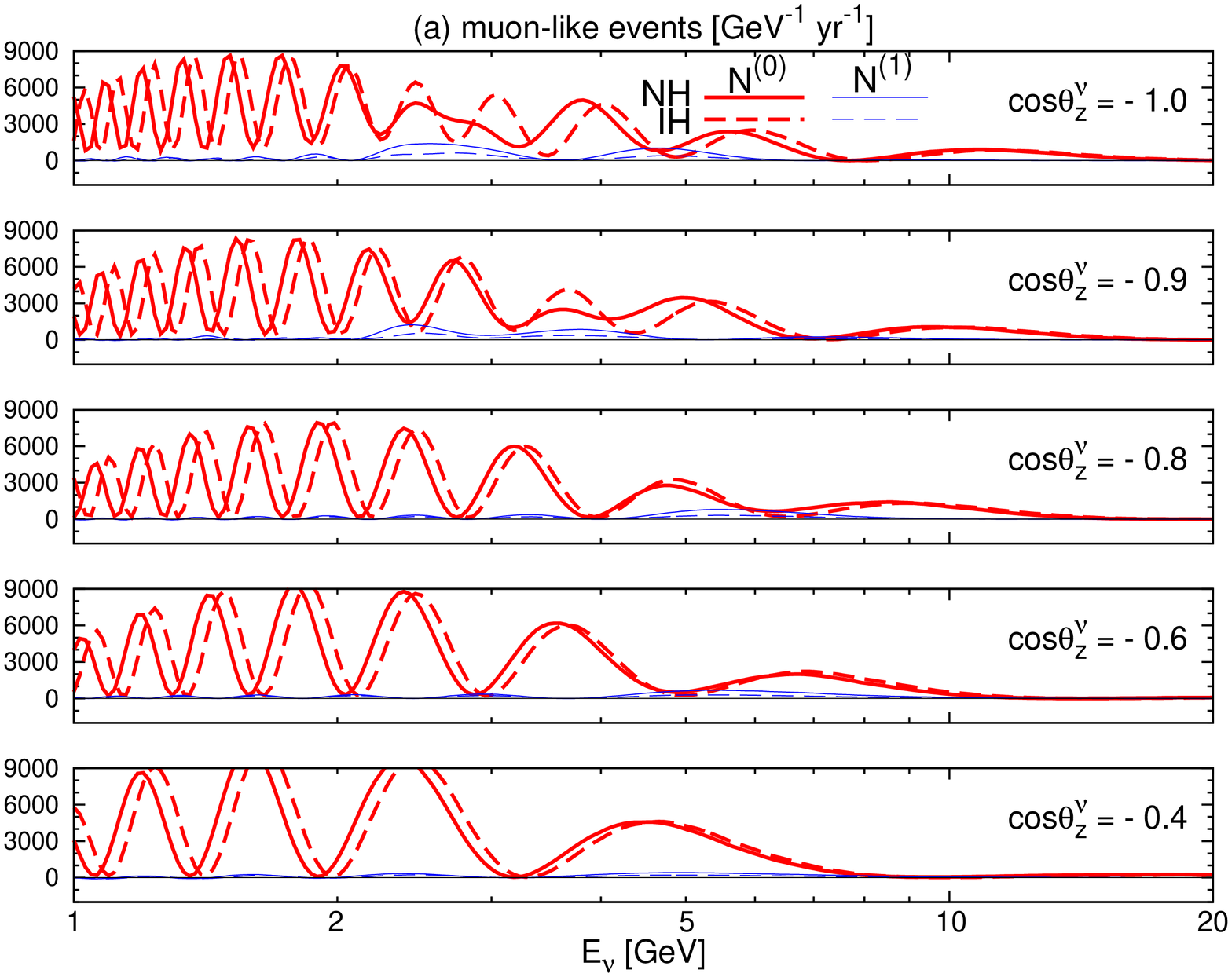}
\includegraphics[height=0.32\textwidth,width=7cm,angle=-90]{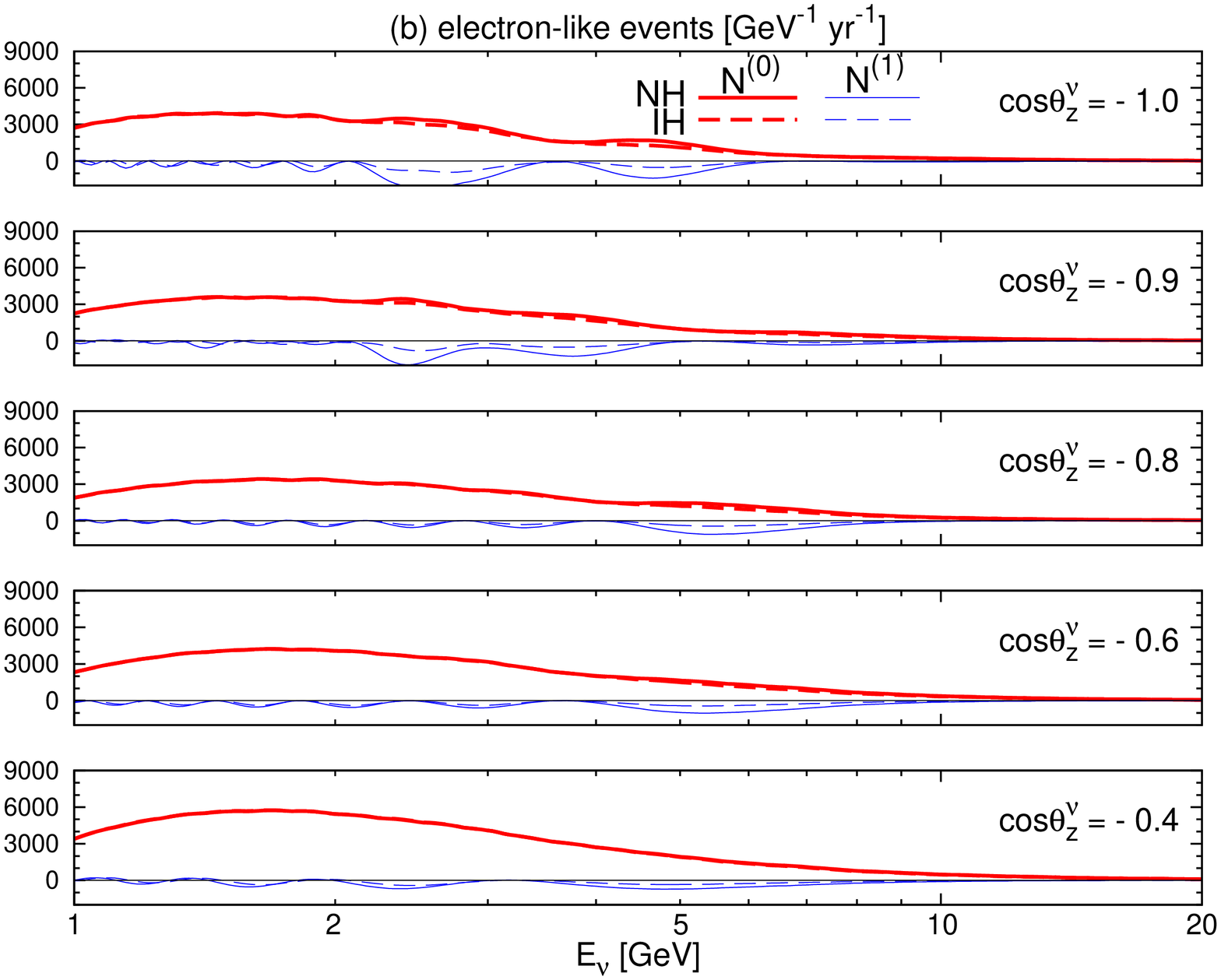}
\includegraphics[height=0.32\textwidth,width=7cm,angle=-90]{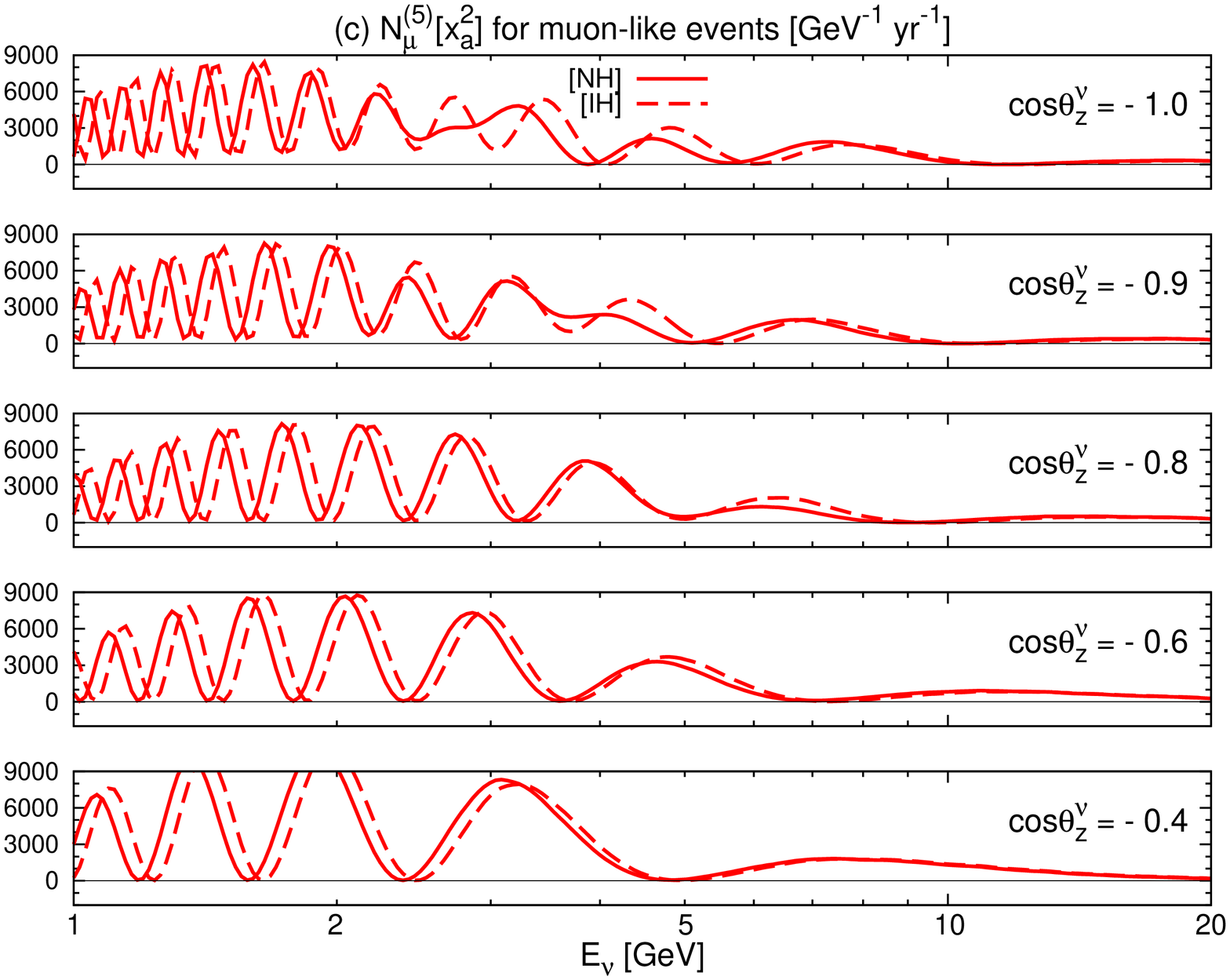}
\caption{The coefficients of $\delta$-independent terms in the propagation-basis
         decomposition of the muon-like and cascade event rates, with 
				 normal hierarchy (NH) [solid curves] and 
				 inverted hierarchy (IH) [dashed curves]. The number of events, 
$N^{(0)}_\alpha + N^{(1)}_\alpha x_{\rm a} + N^{(5)}_\alpha x^2_{\rm a}$,
corresponds to nominal expectation of 40-string PINGU in 1 year.}
\label{fig:N0}
\end{figure}

In \gfig{fig:N0},
we show the relevant coefficients of the $\delta$-independent terms in
(\ref{eq:dNdE}), $N^{(0)}_\mu$ (red curves) and $N^{(1)}_\mu$ (blue curves)
in the left, $N^{(0)}_{\rm e}$ (red curves) and $N^{(1)}_{\rm e}$ (blue curves)
in the center, and $N^{(5)}_\mu$ in the right panels, in the region of 
$1 \, \mbox{GeV} < E_\nu < 20 \, \mbox{GeV}$ for give zenith angles 
$\cos \theta^\nu_{\rm z} = -1, -0.9, -0.8, -0.6, -0.4$ of the neutrino momentum 
direction.
 The nominal number of events per GeV for one-year run of PINGU
is shown along the vertical axis.
 Note that, 
the plots in \gfig{fig:N0} have been updated from those of \cite{Ge2013} by 
accounting for the fiducial volume for 40-string configuration, which is 
obtained from the 20-string configuration \cite{icrc555} by rescaling the 
neutrino energy, $V^{(40)}_{\rm eff}(E_\nu) = V^{(20)}_{\rm eff}(2 \times E_\nu)$.
Since the full-detector simulation of the PINGU detector is not available yet,
and the fact that the tau neutrino contribution is negligible while the 
difference between the effective fiducial volumes for muon and electron 
neutrino is not so large according to a preliminary simulation \cite{Tang11},
we just assume that the effective fiducial volume is the same for different
flavors.

First, for the overall event rates of $N^{(0)}_\alpha$, shown by the red curves
in the left ($\alpha = \mu$) and the center ($\alpha = {\rm e}$) panels, the
cascade event rates have a much smoother energy and angular dependence
than the muon-like event rates. In addition, the cascade event rates
are consistently larger for the normal hierarchy (NH), shown by red-solid 
curves, than for the inverted hierarchy (IH), shown by the red-dashed curves
in the center panel, while the muon-like event rates have strong oscillatory
pattern where the region of NH v.s. IH dominance alter frequently with 
neutrino energy, $E_\nu$, and somewhat also with the zenith angle,
$\cos \theta^\nu_{\rm z}$. These features suggest that the cascade events 
may be more stable against the energy-angular smearing effects than the
muon-like events, as will be discussed in \gsec{sec:scattering} and 
\gsec{sec:resolution}.

Another notable feature in \gfig{fig:N0} is that, $N^{(1)}_\mu$ is always 
positive while $N^{(1)}_{\rm e}$ is always negative, for the coefficients of the 
$x_{\rm a}$ term. More closely examining the curves, we note that the 
difference between NH and IH in the event rates, 
$N^{(0)}_\alpha + N^{(1)}_\alpha \, x_{\rm a}$, tends to increase for the 
negative $x_{\rm a}$, both for the muon-like and cascade events. 
Therefore, we expect the neutrino mass hierarchy discrimination 
by atmospheric neutrino oscillation to be easier for the higher octant,
$x_{\rm a} < 0 (\sin^2 \theta_{\rm a} > 0.5)$, than for the lower octant,
$x_{\rm a} > 0 (\sin^2 \theta_{\rm a} < 0.5)$. 

The coefficient
$N^{(5)}_\mu$ of the quadratic term $x^2_{\rm a}$ is shown in the right panel. 
Note that this quadratic term only appears in the muon-like events.
It can dominate over $N^{(0)}_\mu$ and $N^{(1)}_\mu$ around the oscillation
minima in the
energy range $5 \, \mbox{GeV} \lesssim E_\nu \lesssim 10 \, \mbox{GeV}$ where both
$N^{(0)}_\mu$ and $N^{(1)}_\mu$ are tiny but $N^{(5)}_\mu$ is larger by an
order of magnitude, because the oscillation phases (the location of this 
minima) are different between $N^{(0)}_\mu$ and $N^{(5)}_\mu$ \cite{Ge2013}. 
In the whole range, $N^{(5)}_\mu$ is always positive. More importantly,
$N^{(5)}_\mu$ has opposite phase w.r.t. $N^{(0)}_\mu$. For example,
$N^{(5)}_\mu$ peaks around $E_\nu = 5 \, \mbox{GeV}$ and $\cos \theta^\nu_{\rm z}=-1$,
with larger event rate for IH, where it is a minimum for $N^{(0)}_\mu$ with
larger event rate for NH. The hierarchy sensitivity in $N^{(5)}_\mu$ always 
cancels with the one in $N^{(0)}_\mu$. In other words, the quadratic term 
reduces the hierarchy sensitivity in the muon-like event rates. 
Summing up, the mass hierarchy dependence of the 
event rate, $N^{(0)}_\mu + N^{(1)}_\mu \, x_{\rm a} + N^{(5)}_\mu \, x^2_{\rm a}$,
tends to decrease as $x^2_{\rm a}$ increases or $x_{\rm a}$ increases.
Consequently, it decreases when $x_{\rm a}$ 
is positive and $x^2_{\rm a}$ increases. If $x_{\rm a}$ is negative, the 
trend depends on which one of the effects from the linear term and
quadratic term of $x_{\rm a}$ dominates. Since $N^{(5)}_\mu$ can dominate
over $N^{(0)}_\mu$ and $N^{(1)}_\mu$ when $x_{\rm a} \approx -0.2$, the 
effect of the quadratic term dominates. Hence, the sensitivity decreases
when $x_{\rm a}$ is negative and decreases.

Due to the Earth
matter potential, neutrinos travelling through the mantle can experience MSW 
resonance \cite{msw1,msw2,msw3,msw4} while those travelling through the core 
can experience an extra parametric resonance 
\cite{parametric1,parametric2,parametric3,parametric4} which is also known as
oscillation-length resonance \cite{length1,length2,length3,length4}. This only 
happens for the case of neutrino with NH or antineutrino 
with IH. If neutrino events can be distinguished from antineutrino events, 
the sensitivity to the neutrino mass hierarchy by observing the 
atmospheric neutrino oscillation pattern can improve significantly.
The study in \cite{Ge2013}, as summarized above in \gfig{fig:N0}, assumes that
PINGU cannot discriminate between neutrino and antineutrino events, and hence
all the hierarchy dependence of the observed event numbers are due to the
difference in the neutrino and antineutrino fluxes \cite{honda12}, and in
the CC cross sections \cite{neugen}. Luckily, both the fluxes and the cross
sections are larger for neutrino than antineutrino, resulting in the 
significant hierarchy dependence for muon-like and cascade events.
In the following \gsec{sec:scattering}, we will show how one can use the 
inelasticity distribution of muon CC events to discriminate between $\nu_\mu$
and $\bar \nu_\mu$ events, which can further improve the hierarchy 
sensitivity of the experiment.

\section{Neutrino Scattering and Reconstruction Procedures}
\label{sec:scattering}

Experimentally, neutrino momentum cannot be directly measured. 
Neutrinos interact with the 
target, generating various final-state particles, some of which can leave traces 
in the detector. These traces are measured to reconstruct the energy and momentum 
of the final-state particles and then the incident neutrino momentum is 
estimated via the energy-momentum conservation. 
Because the detector responses to the muon, electron, and hadron are different,
reconstruction of the incident neutrino momentum is not only difficult but also
depends on the reaction. The visible neutrino energy, $E_{\rm vis}$,
and its visible momentum direction, 
$\cos \theta^{\rm vis}_{\rm z}$, distribute around
their true values, $E_\nu$ and $\cos \theta^\nu_{\rm z}$. 
The resulting observable distributions significantly 
smear the oscillation patterns in neutrino event rates, as shown in \gfig{fig:N0}, 
and reduce the sensitivities discussed in \cite{Ge2013}. On the brighter side, 
the topology of the final-state particles in the charged current (CC) scattering 
of neutrino is significantly different from that of antineutrino. This 
feature can be used to distinguish neutrino from antineutrino statistically
for CC events and 
enhance the sensitivity to the neutrino mass hierarchy.

\subsection{Neutrino Scattering and Inelasticity Distribution}
\label{sec:y}

Since the information of neutrino oscillation is kept only in the CC scattering, 
we will focus on the following processes,
\begin{subequations}
\begin{eqnarray}
  \nu_\ell + N
& \rightarrow &
  \ell + X \,,
\\
  \bar \nu_\ell + N
& \rightarrow &
  \bar \ell + X \,,
\end{eqnarray}
\end{subequations}
where $N$ is the target nucleon and $X$ represents the final-state hadrons, 
for $\ell = e \mbox{ or } \mu$.
In terms of the Bjorken scaling variables, the differential 
cross section per nucleon can be expressed in the parton model as 
\cite{Gandhi95,Gandhi98},
\begin{subequations}
\begin{eqnarray}
  \frac {\partial^2 \sigma_\nu}{\partial x \partial y}
& = &
  \frac {2 G^2_F M E_\nu} \pi
\left[
  x q(x, Q^2)
+ x \bar q(x, Q^2) (1 - y)^2
\right] \,,
\label{eq:cs1nu}
\\
  \frac {\partial^2 \sigma_{\bar \nu}}{\partial x \partial y}
& = &
  \frac {2 G^2_F M E_{\bar \nu}} \pi
\left[
  x \bar q(x, Q^2)
+ x q(x, Q^2) (1 - y)^2
\right] \,,
\label{eq:cs1nubar}
\end{eqnarray}
\label{eq:cs1}
\end{subequations}
\hspace{-2.5mm}
with $x \equiv Q^2/2 M (E_\nu - E_\ell)$ and inelasticity 
$y \equiv (E_\nu - E_\ell) / E_\nu$ for neutrino and similarly for antineutrino. 
Here $q(x, Q^2)$ and $\bar q(x,Q^2)$
are the quark and antiquark distributions in the nucleon measured at the 
momentum transfer scale of $Q^2 \equiv 2 E_\nu E_\ell (1 - \cos \theta_\ell)$.
Because the quark distribution $q(x, Q^2)$ is much larger than the antiquark 
distribution $\bar q(x, Q^2)$ in nuclei, the $\nu$--CC events are expected to 
have relatively flat distribution of inelasticity, whereas the $\bar \nu$--CC
events have strong suppression at large inelasticity due to the $(1-y)^2$
factor multiplying the quark distribution $q(x,Q^2)$ in (\ref{eq:cs1nubar}).
After integrating over $y$, the total cross section of antineutrino is
roughly one third of that of neutrino for an isoscalar nucleon. We therefore
expect that at small inelasticity, $1 - y \approx 1$, the antineutrino cross
section $\sigma_{\bar \nu}$ is as large as the neutrino cross section
$\sigma_\nu$, but at large inelasticity, $1 - y \approx 0$, $\sigma_{\bar \nu}$
is much smaller than $\sigma_\nu$. We can enhance the purity of the
$\nu$--CC events by selecting those events with $1 - y \approx 0$, where those
events at $1 - y \approx 1$ is a mixture of $\nu$-- and $\bar \nu$--CC
events, whose ratio may reflect that of the original fluxes.

\begin{figure}[h]
\centering
\includegraphics[height=12cm,width=7cm,angle=-90]{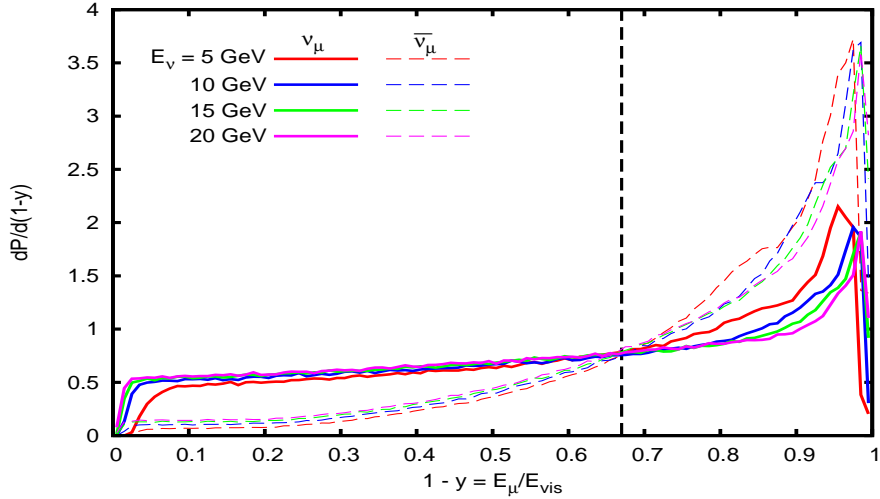}
\caption{Inelasticity distribution of neutrino (solid curves) and antineutrino 
(dashed curves) CC scattering.}
\label{fig:inelasticity}
\end{figure}

We use GENIE \cite{genie} to obtain the inelasticity distribution as shown
in \gfig{fig:inelasticity}, for $\nu_\mu$-- and $\bar \nu_\mu$--CC events
off water target. The normalized cross sections are shown by solid curves
for $\nu_\mu$ and by dashed curves for $\bar \nu_\mu$, for neutrino energies
of 5, 10, 15, and $20 \, \mbox{GeV}$. It is amusing to note that the 
normalized distributions cross at $1 - y \approx 0.67$, which is stable in
the energy range of $5 \, \mbox{GeV} \lesssim E_\nu \lesssim 20 \, \mbox{GeV}$. 
Note that the inelasticity shown in \gfig{fig:inelasticity} is defined in terms 
of the visible energy $E_{\rm vis}$, instead of $E_\nu$,
\begin{equation}
  1 - y
\equiv
  \frac {E_\ell}{E_{\rm vis}} \,,
\label{eq:y}
\end{equation}
where $E_{\rm vis}$ will be defined in the next section.
The smearing effects detailed in \gsec{sec:reconstruction} and 
\gsec{sec:resolution} have already been accounted for in order to make a 
realistic illustration.

Experimentally, there are several ways of distinguishing neutrino from 
antineutrino. The charge of the lepton can be measured by applying magnetic field
to the detector, such as ICAL \cite{ical1} and a liquid argon detector 
\cite{liquid1}, to tell if it is produced in CC scattering by neutrino or 
antineutrino. For those detectors
implementing magnetic field is not possible, such as Super-K \cite{sk1,sk2}, 
distinct signal, such as the delayed signal in single-ring electron events, or 
characteristic distribution, such as the inelasticity dependence of multi-ring 
electron events, can be used. The latter method can also apply to PINGU
\cite{Ribordy12,Ribordy13} for muon-CC events, which will be studied in 
the following sections.

\subsection{Reconstruction Procedures}
\label{sec:reconstruction}

Neutrino scattering can smear the oscillation pattern since it is impossible to 
directly measure the neutrino momentum. The information can only be partially 
retrieved by 
reconstructing from the final-state particles that are observable. For CC events, 
in principle it is possible to reconstruct the incident neutrino momentum, 
whereas a sizable fraction of the momentum is lost in the neutral current (NC) 
as well $\tau$-CC events. This effect is especially severe for atmospheric 
neutrino experiments where the direction of the incident neutrino is unknown. 
In this section, we discuss the energy and zenith angle reconstruction procedures,
leaving the detector resolutions to be studied in \gsec{sec:resolution}.
The reconstruction procedures assumed here are intuitive and need to be 
enriched by full-detector simulation which is not available yet.

\subsubsection{Energy Reconstruction}
\label{sec:Ereconst}

PINGU measures the Cherenkov light from the final-state particles by recording
their energy and arriving time. These information can be used to reconstruct 
the energy and momentum of the final-state particles \cite{reconstruction}.

Depending on the particle species, the Cherenkov light yield is different. For 
charged lepton, muon or electron, its energy can be estimated from the Cherenkov 
light radiation, while the yield is significantly lower for other particle 
that can induce hadronic cascades, which consist of electromagnetic showers 
from $\pi^0$'s and radiations from charged pions and nucleon excitations. 
In the following study, we assign the equivalent visible energy of a hadronic 
cascade to be 80\% of that of an electromagnetic shower \cite{yield1,yield2,yield3}. 
Then, the cascade Cherenkov light energy 
can be expressed as,
\begin{eqnarray}
  E_{\rm cas}
\equiv
  0.8 \times (E_\nu - E_{\nu '} - E_\ell) + E_{\rm e} \,,
\end{eqnarray}
where $E_\nu$ and $E_{\nu '}$ are the energies of the incident and the final-state
neutrinos, respectively. For $\mu$--CC and e-CC, $E_{\nu '} = 0$ and it can be 
large for NC and $\tau$-CC events. For $\mu$--CC events, $E_\ell = E_\mu$ and
$E_{\rm e} = 0$, while for e--CC events, $E_\ell = E_{\rm e}$. 

There are two typical topologies on PINGU. Muon with large enough energy,
$E_\mu > 1 \, \mbox{GeV}$, leaves a clear track due to its long lifetime while 
the cascade produced by other particles has a spherical structure. Because 
of this difference, the energies of muon and cascade can be
reconstructed independently. This distinguishability makes estimating the 
muon inelasticity possible, as defined in (\ref{eq:y}). On the other hand, 
the electron shower cannot be distinguished from cascade, hence,
its energy cannot be measured separately. For muon with small energy,
$E_\mu < 1 \, \mbox{GeV}$, the track may not be long and clear enough to be 
identified, and it is counted as cascade events. In addition, 10\% 
of the energetic muons, $E_\mu > 1 \, \mbox{GeV}$, are assumed to be 
mis-identified as cascade events \cite{Winter13}. For $\tau$, it decays
very quickly into muon, electron, or hadrons. For all these cases, the visible 
energy can then be estimated as,
\begin{equation}
  E_{\rm vis}
\equiv
  E_\mu + \frac {E_{\rm cas}}{0.8}
=
\begin{cases}
  E_\nu  & \mu-CC \, (E_\mu > 1 \, \mbox{GeV with 90\% } \mu\mbox{-ID}) \,, \\
  E_\nu + 0.25 E_{\ell'} & 
  \begin{cases}
    e-CC \,,  \\[-1mm]
    \mu-CC \, (E_\mu < 1 \, \mbox{GeV}, E_\mu > 1 \, \mbox{GeV with 10\% } \mu\mbox{-misID}) \,,
  \end{cases}  \\[3mm]
  E_\nu - E_{\nu '} & 
  \begin{cases}
    \tau-CC \mbox{ with } \tau \rightarrow \mu \, (E_\mu > 1 \, \mbox{GeV with 90\% } \mu\mbox{-ID}) \,, \\[-1mm]
		\tau-CC \mbox{ with } \tau \rightarrow \mbox{hadrons} \,, \\[-1mm]
    NC \,,
  \end{cases} \\[1mm]
  E_\nu - E_{\nu '} + 0.25 E_{\ell'} & \tau-CC  
  \begin{cases} 
    \tau \rightarrow e \,, \\[-1mm]
    \tau \rightarrow \mu \, (E_\mu < 1 \, \mbox{GeV}, E_\mu > 1 \, \mbox{GeV with 10\% } \mu\mbox{-misID}) \,,
  \end{cases}
\end{cases}
\label{eq:Evis}
\end{equation}
where $\ell'$ denotes the electron or the mis-identified muon. 
In this definition, $E_{\rm vis}$ is the visible energy after rescaling the
cascade energy $E_{\rm cas}$ and adding it to the muon energy $E_\mu$, so that
for muon-like events $E_{\rm vis}$ is exactly the neutrino energy.

For e--CC events, all the final-state energies are recorded as 
shower energies, and hence the electron energy $E_{\rm e}$ is 25\% 
overestimated. For $\mu$--CC events, if the muon is mis-identified as a 
shower, or if the muon energy $E_\mu$ is too low for the track reconstruction, 
its energy is again overestimated by 25\%. The visible energy $E_{\rm vis}$ 
is larger than the neutrino energy $E_\nu$, with the difference depending on 
$E_{\ell'}$ (the energy of final state electron or that of final state muon
when it is mis-identified as electron), 
and hence contributes to the smearing effect during energy 
reconstruction. The typical size of the final-state lepton energy ($E_{\ell '}$)
distribution, divided by a factor of $4$, is therefore the characteristic 
energy resolution from energy reconstruction. Since the inelasticity distribution
is quite stable as shown in \gfig{fig:inelasticity}, this characteristic
energy resolution is proportional to $E_\nu$, independent of the neutrino energy. 
On average, the lepton takes away about 60\% of neutrino and 80\% of antineutrino
energies. 
The smearing effect in energy reconstruction can hence be as large as
$\sigma_{\rm E} \approx 0.15 \times (E_\nu / \mbox{GeV})$ for neutrino and 
$\sigma_{\rm E} \approx 0.2 \times (E_\nu / \mbox{GeV})$ for antineutrino. 
This explains the linear dependence of the energy resolution on the neutrino 
energy, $\sigma_{\rm E} \approx 0.25 \times (E_\nu / \mbox{GeV})$, as
found in \cite{icrc555}. The discrepancy may be due to the energy
resolution originating from statistical fluctuation,
$\sigma_{\rm E} \approx 0.2 \times \sqrt{E_\nu / \mbox{GeV}}$, 
as will be discussed in \gsec{sec:Eresol}. It is worth noting here that our 
very naive treatment of the energy reconstruction error tends to reproduce the 
simulation results \cite{icrc555} when combined with the expected statistical
error of the measurement. The above estimation applies only to e-CC 
events. For NC and $\tau$-CC events with $\tau$ decaying into hadrons, 
the visible energy $E_{\rm vis}$ is much smaller than the neutrino energy $E_\nu$ 
due to the energy $E_{\nu'}$ of the final-state neutrinos. Therefore, the smearing effect due to energy 
reconstruction is proportional to the distribution of the final-state neutrino 
energy $E_{\nu '}$ which is also expected to be proportional to the neutrino 
energy. For the fourth case in (\ref{eq:Evis}), $\tau$--CC with tau decaying
into an electron or a mis-identified muon, the smearing effect comes from
both the final-state neutrino and lepton.

\subsubsection{Zenith Angle Reconstruction}
\label{sec:Zrec}

Depending on the path, which is a function of the zenith angle, atmospheric 
neutrino experiences different matter potential and baseline length. It is 
necessary to recover the zenith angle of the incident neutrino in order to 
measure the neutrino oscillation pattern. This can be achieved by reconstructing
the momentum of the incident neutrino.

Since muon has a long track in the PINGU detector, its direction can be
determined with a much higher precision than the cascade. If 
there is a muon in the final state, such as the $\mu$-CC channel and the 
$\tau$-CC channel with $\tau$ decaying into an energetic muon, 
$E_\mu > 1 \, \mbox{GeV}$, the reconstructed direction is mainly determined 
by the momentum of this muon. The same scenario also applies to the 
case with an energetic electron which can produce a forward radiation
whose direction dominates in the cascade radiation.
If the energy carried away by lepton is not large enough, the Cherenkov light
from lepton will be overwhelmed by the hadronic radiation. Then the total visible 
momentum can only be estimated from the total visible momentum in the final state.
These two topologies can be expressed as,
\begin{equation}
  \vec P_{\rm vis}
=
\begin{cases}
  \vec P_\ell & \hspace{3mm} CC \, (E_\ell > 1 \, \mbox{GeV}) \,, \\[1mm]
  \vec P_\nu - \vec P_{\nu'} & 
  \begin{cases}
    CC \, (E_\ell < 1 \, \mbox{GeV}) \,, \\[-1mm]
    NC \,,
  \end{cases}
\end{cases}
\end{equation}
where $\ell$ stands for muon or electron, $\vec P_{\nu'}$ denotes the vector 
sum of all the final-state neutrinos in NC and $\tau$-CC events, whereas
$\vec P_{\nu'} = 0$ for muon-like and cascade CC events. 
In both cases, the momentum 
of the incident neutrino cannot be exactly reconstructed. 
For CC with $E_\ell > 1 \, \mbox{GeV}$, in principle, the direction 
of the hadronic cascade can also be used as supplementary information to
help estimating the neutrino direction. But it has much worse angular resolution
and we do not attempt to reconstruct the hadronic momentum vector for events
with $E_\ell > 1 \, \mbox{GeV}$.

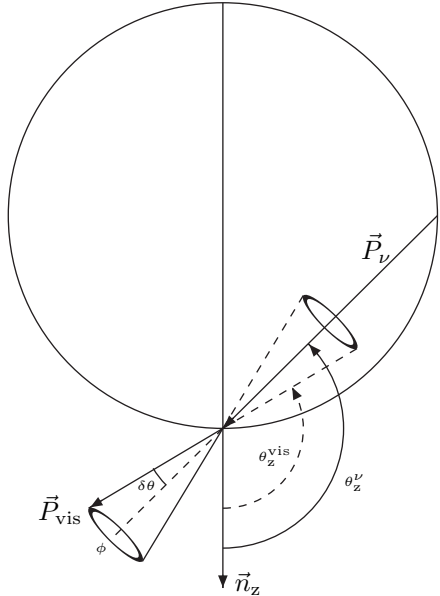
\begin{figure}[h]
\centering
\begin{picture}(300,210)(0,90)
\CArc(150,220)(80,0,360)
\Line(150,300)(150,80)
\ArrowLine(150,85)(150,80)
\Text(160,82)[]{$\vec n_{\rm z}$}
\Line(230,220)(150,140)
\ArrowLine(154,144)(150,140)
\DashLine(150,140)(110,100){3}
\Text(105,95)[]{\tiny $\phi$}
\Text(208,203)[b]{$\vec P_\nu$}
\Line(150,140)(100,110)
\ArrowLine(105,113)(100,110)
\Line(150,140)(120,90)
\DashLine(150,140)(200,170){3}
\DashLine(150,140)(180,190){3}
\Text(98,110)[r]{$\vec P_{\rm vis}$}
\Oval(110,100)(3,13)(135)
\CArc(150,140)(30,210,225)
\Text(122,119)[]{\tiny $\delta \theta$}
\Oval(190,180)(3,13)(135)
\CArc(150,140)(45,270,45)
\ArrowArc(150,140)(45,38,45)
\Text(200,120)[]{\tiny $\theta^\nu_{\rm z}$}
\DashCArc(150,140)(30,270,30){3}
\DashArrowArc(150,140)(30,23,30){3}
\Text(170,130)[]{\tiny $\theta^{\rm vis}_{\rm z}$}
\end{picture}
\caption{Kinematics of neutrino scattering.}
\label{fig:scattering}
\end{figure}

For convenience, the scattering process has been illustracted in 
\gfig{fig:scattering}, where $\theta^\nu_{\rm z}$ and $\theta^{\rm vis}_{\rm z}$ 
are the zenith angles of the incident neutrino momentum $\vec P_\nu$ and the 
visible momentum $\vec P_{\rm vis}$, respectively, parametrized as,
\begin{equation}
- \vec P_\nu
\equiv
  \left| \vec P_\nu \right|
\left\lgroup
\begin{matrix}
  \sin \theta^\nu_{\rm z} \\
  0 \\
  \cos \theta^\nu_{\rm z}
\end{matrix}
\right\rgroup \,,
\qquad
- \vec P_{\rm vis}
\equiv
  \left| \vec P_{\rm vis} \right|
\left\lgroup
\begin{matrix}
  \sin \theta^{\rm vis}_{\rm z} \cos \phi_{\rm vis} \\
  \sin \theta^{\rm vis}_{\rm z} \sin \phi_{\rm vis} \\
  \cos \theta^{\rm vis}_{\rm z}
\end{matrix}
\right\rgroup \,.
\end{equation}
The opening angle between $\vec P_\nu$ and $\vec P_{\rm vis}$ is denoted 
as $\delta \theta$ while $\phi$ is the azimuthal angle of $\vec P_{\rm vis}$ 
with respect to $\vec P_\nu$ in the neutrino frame, where $\vec P_\nu$ gives 
the polar axis and the azimuthal angle is measured from the plane which
contains the zenith direction vector $\vec n_{\rm z}$ at the South Pole, as shown in 
\gfig{fig:scattering}. With a given zenith angle 
$\theta^\nu_{\rm z}$ of the incident neutrino, the visible zenith angle 
$\theta^{\rm vis}_{\rm z}$ is a function of $\delta \theta$ and $\phi$,
\begin{equation}
  \cos \theta^{\rm vis}_{\rm z}
=
  \cos \theta^\nu_{\rm z} \cos \delta \theta 
- \sin \theta^\nu_{\rm z} \sin \delta \theta \cos \phi \,.
\label{eq:tvis}
\end{equation}
From this, we can observe that the visible zenith angle takes a value in the 
range of,
\begin{equation}
  \min (0, \theta^\nu_{\rm z} - \delta \theta)
\leq 
  \theta^{\rm vis}_{\rm z}
\leq 
  \max (\pi, \theta^\nu_{\rm z} - \delta \theta) \,,
\end{equation}
which is always within the defined range $[0,\pi]$ of the zenith angle.
At very high energies, the opening angle $\delta \theta$ is tiny, and the
approximation $\theta^{\rm vis}_{\rm z} \approx \theta^\nu_{\rm z}$ holds.
At energies below $10 \, \mbox{GeV}$, the opening angle can be significant,
and the projection from $\theta^\nu_{\rm z}$ to $\theta^{\rm vis}_{\rm z}$
needs to be done carefully as follows.

The distribution of $\vec P_{\rm vis}$ is solely determined by the neutrino 
interactions. In the neutrino frame, the event distribution after scattering 
depends on the polar angle 
$\delta \theta$, but not on the azimuthal angle $\phi$ or
the neutrino zenith angle $\theta^\nu_{\rm z}$,
\begin{equation}
\left.
  \frac {\partial^3 N}
				{\partial E_{\rm vis} \partial \delta \theta \partial \phi}
\right|_{E_\nu}
\equiv
  \frac 1 {2 \pi}
  \mathbb T(E_\nu | E_{\rm vis}, \delta \theta) \,,
\end{equation}
where $N$ represents the neutrino event number and the transfer table $\mathbb T$ 
describes the distribution of $\vec P_{\rm vis}$ for a given neutrino energy
$E_\nu$. In the current study, we use GENIE \cite{genie} to generate the transfer
table $\mathbb T(E_\nu | E_{\rm vis}, \delta \theta)$. This universal transfer pattern
needs to be projected onto the earth frame in which the zenith angle is measured, 
as illustrated in \gfig{fig:scattering}. Since the azimuthal angle $\phi$ varies
freely, the observed zenith angle $\theta^{\rm vis}_{\rm z}$ distributes around 
the neutrino zenith angle $\theta^\nu_{\rm z}$,
\begin{equation}
\left. \int
  \frac {\partial^3 N}
				{\partial E_{\rm vis} \partial \delta \theta \partial \phi}
\right|_{E_\nu}
\hspace{-3mm}
  \delta (
  \cos \theta^{\rm vis}_{\rm z}
-
  \cos \theta^\nu_{\rm z} \cos \delta \theta 
+ \sin \theta^\nu_{\rm z} \sin \delta \theta \cos \phi ) d \phi d \delta \theta
\left.
=
  \frac {\partial^2 N}
				{\partial E_{\rm vis} \partial \cos \theta^{\rm vis}_{\rm z}}
\right|_{E_\nu, \theta^\nu_{\rm z}}
\hspace{-3mm}
\equiv
  \mathbb T(E_\nu, \theta^\nu_{\rm z} | E_{\rm vis}, \theta^{\rm vis}_{\rm z}) \,,
\end{equation}
by using the relation (\ref{eq:tvis}).
The integration over $\phi$ gives the Jacobian, and we find,
\begin{equation}
  \mathbb T(E_\nu, \theta^\nu_{\rm z} | E_{\rm vis}, \theta^{\rm vis}_{\rm z})
=
  \int
  \frac 1 {2 \pi}
  \frac {\mathbb T(E_\nu | E_{\rm vis}, \delta \theta) d \delta \theta}
  			{|\sin \theta^\nu_{\rm z} \sin \delta \theta \sin \phi|}
=
  \int
  \frac 1 {2 \pi}
  \frac {\mathbb T(E_\nu | E_{\rm vis}, \delta \theta) d \delta \theta}
				{\sqrt{\sin^2 \theta^\nu_{\rm z} \sin^2 \delta \theta
             -(\cos \theta^\nu_{\rm z} \cos \delta \theta - \cos \theta^{\rm vis}_{\rm z})^2}} \,.
\label{eq:Tpprime}
\end{equation}
Now the transfer table $\mathbb T(E_\nu, \theta^\nu_{\rm z} | E_{\rm vis}, \theta^{\rm vis}_{\rm z})$ is measured in the earth frame and
gives the probability distribution of $\vec P_{\rm vis}$ as a function 
of the visible energy $E_{\rm vis}$ and the visible zenith angle 
$\theta^{\rm vis}_{\rm z}$, given neutrino energy $E_\nu$ and zenith angle
$\theta^\nu_{\rm z}$. The visible event rate distribution can be obtained by
convoluting the neutrino event rates with $\mathbb T(E_\nu, \theta^\nu_{\rm z} | E_{\rm vis}, \theta^{\rm vis}_{\rm z})$,
\begin{equation}
  \frac {\partial^2 N(E_{\rm vis}, \theta^{\rm vis}_{\rm z})}
				{\partial E_{\rm vis} \partial \cos \theta^{\rm vis}_{\rm z}}
=
  \int
  \frac {\partial^2 N(E_\nu, \theta^\nu_{\rm z})}
				{\partial E_\nu \partial \cos \theta^\nu_{\rm z}}
  \mathbb T(E_\nu, \theta^\nu_{\rm z} | E_{\rm vis}, \theta^{\rm vis}_{\rm z})
  d E_\nu d \cos \theta^\nu_{\rm z} \,.
\label{eq:T}
\end{equation}

\subsection{Event Rates Smeared by Neutrino Scattering and Reconstruction Procedures}

\begin{figure}[h!]
\centering
\includegraphics[height=0.32\textwidth,width=7cm,angle=-90]{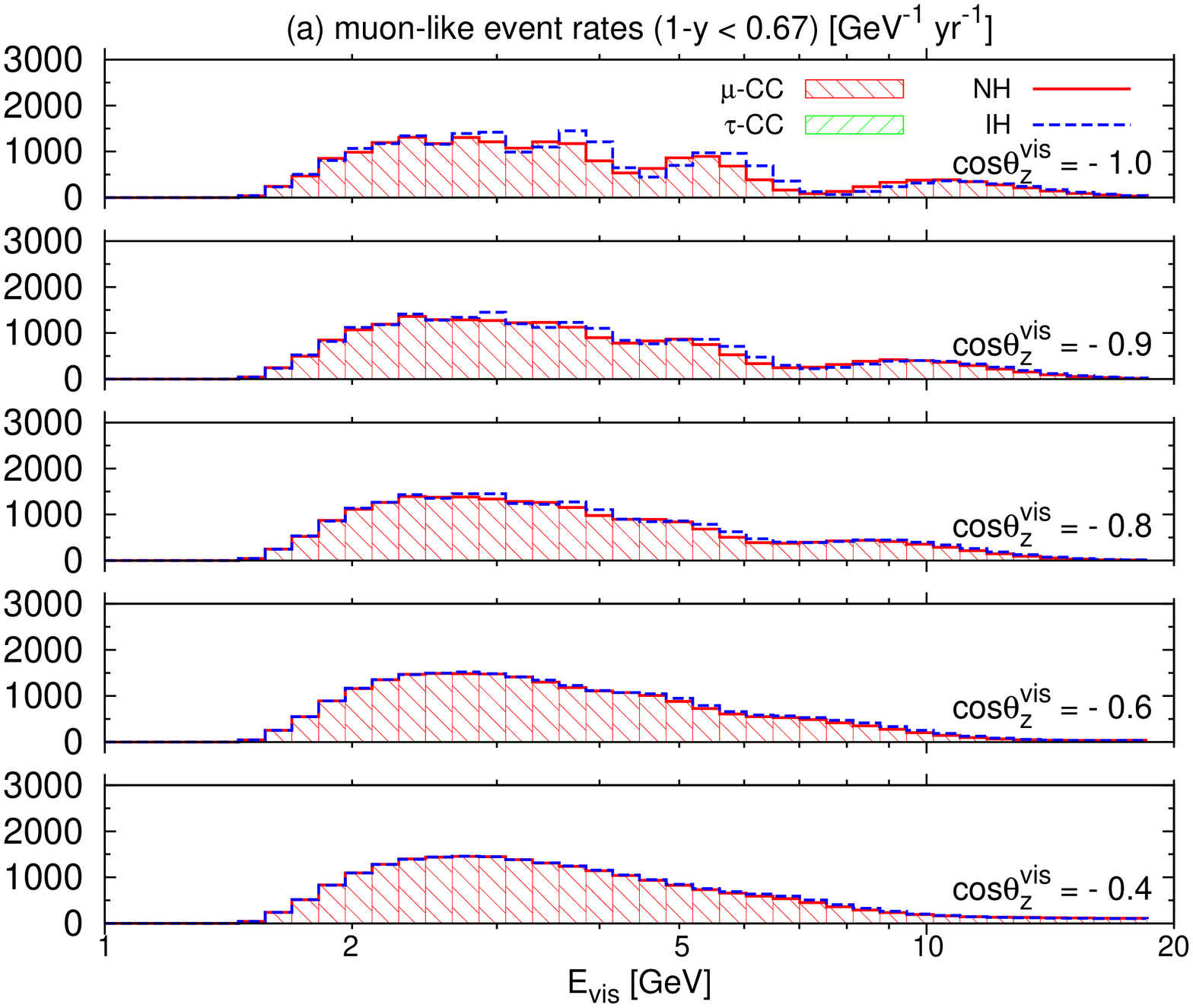}
\includegraphics[height=0.32\textwidth,width=7cm,angle=-90]{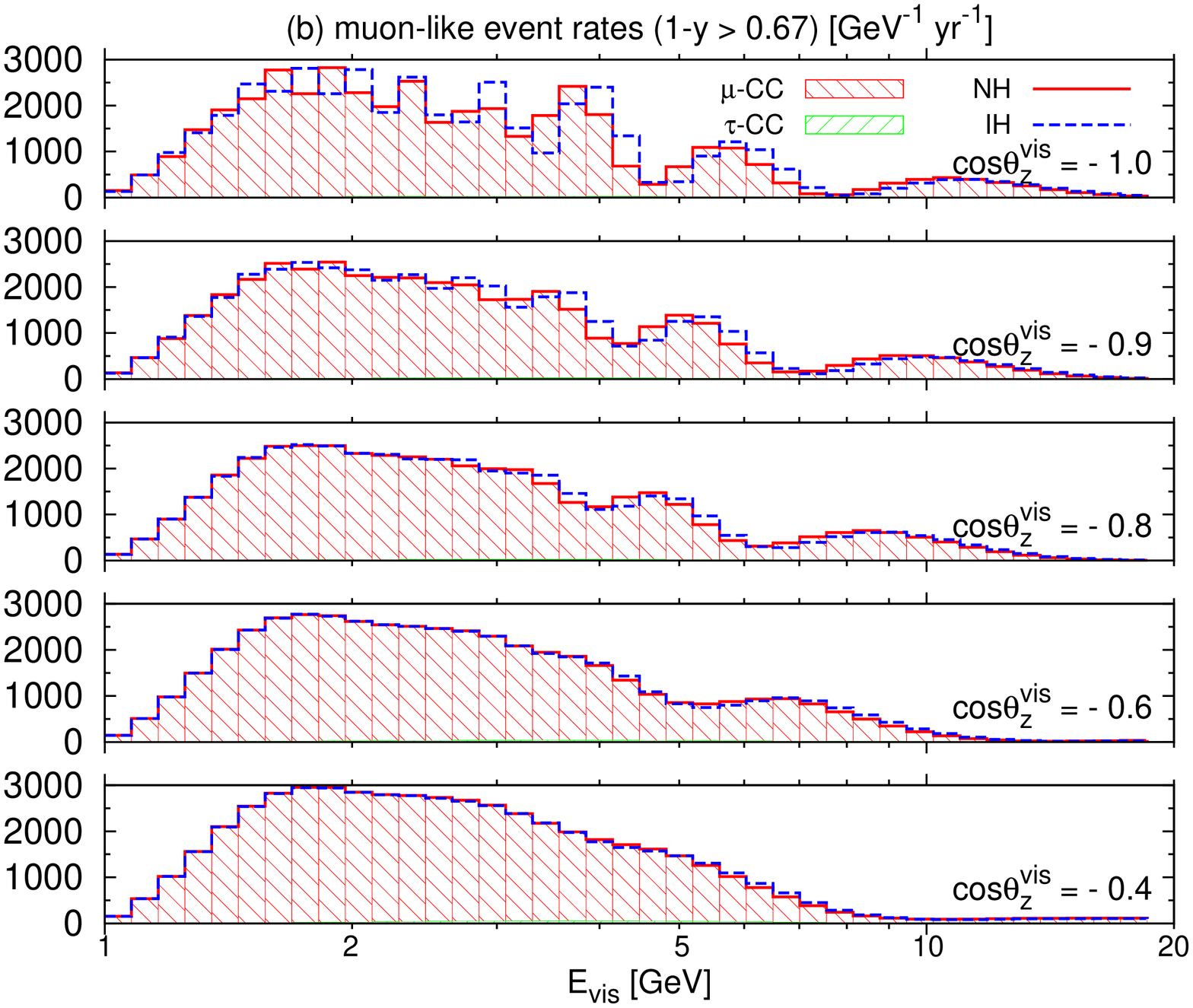}
\includegraphics[height=0.32\textwidth,width=7cm,angle=-90]{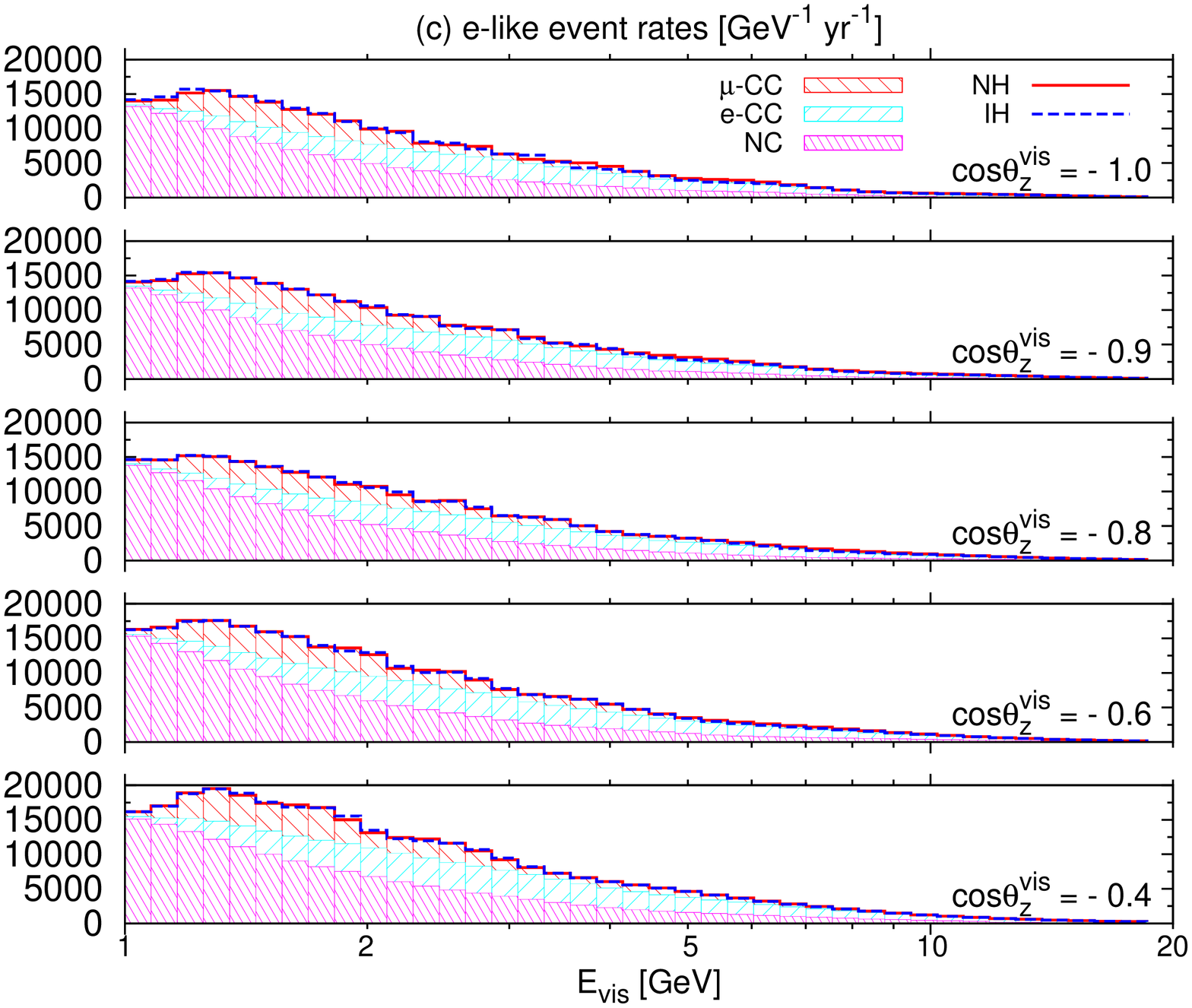}
\vspace{5mm}
\caption{Event rates, smeared by neutrino scattering together with energy and zenith angle 
				 reconstruction procedures, of (a) muon-like channel for $1 - y < 0.67$, 
				 (b) muon-like channel for $1 - y > 0.67$, and (c) cascade channel,
         with NH (red-solid curves) and IH (blue dashed curves) , in 1-year 
				 run of PINGU.}
\label{fig:N1}
\end{figure}

In \gfig{fig:N1}, we show the event rates smeared by neutrino scattering and 
reconstruction procedures in \gsec{sec:reconstruction}, for both muon- and 
cascade channels. For the muon-like events in the panels (a) and (b), 
there are two sources, one from $\mu$-CC and the other from $\tau$-CC with $\tau$
decaying to $\mu$, with muon energy $E_\mu > 1 \, \mbox{GeV}$ as defined 
in (\ref{eq:Evis}). Those events that do not have a muon or the muon is 
mis-identified
are all classified as cascade events in the right panel (c).

\begin{figure}[h!]
\centering
\vspace{4mm}
\includegraphics[height=0.32\textwidth,width=7cm,angle=-90]{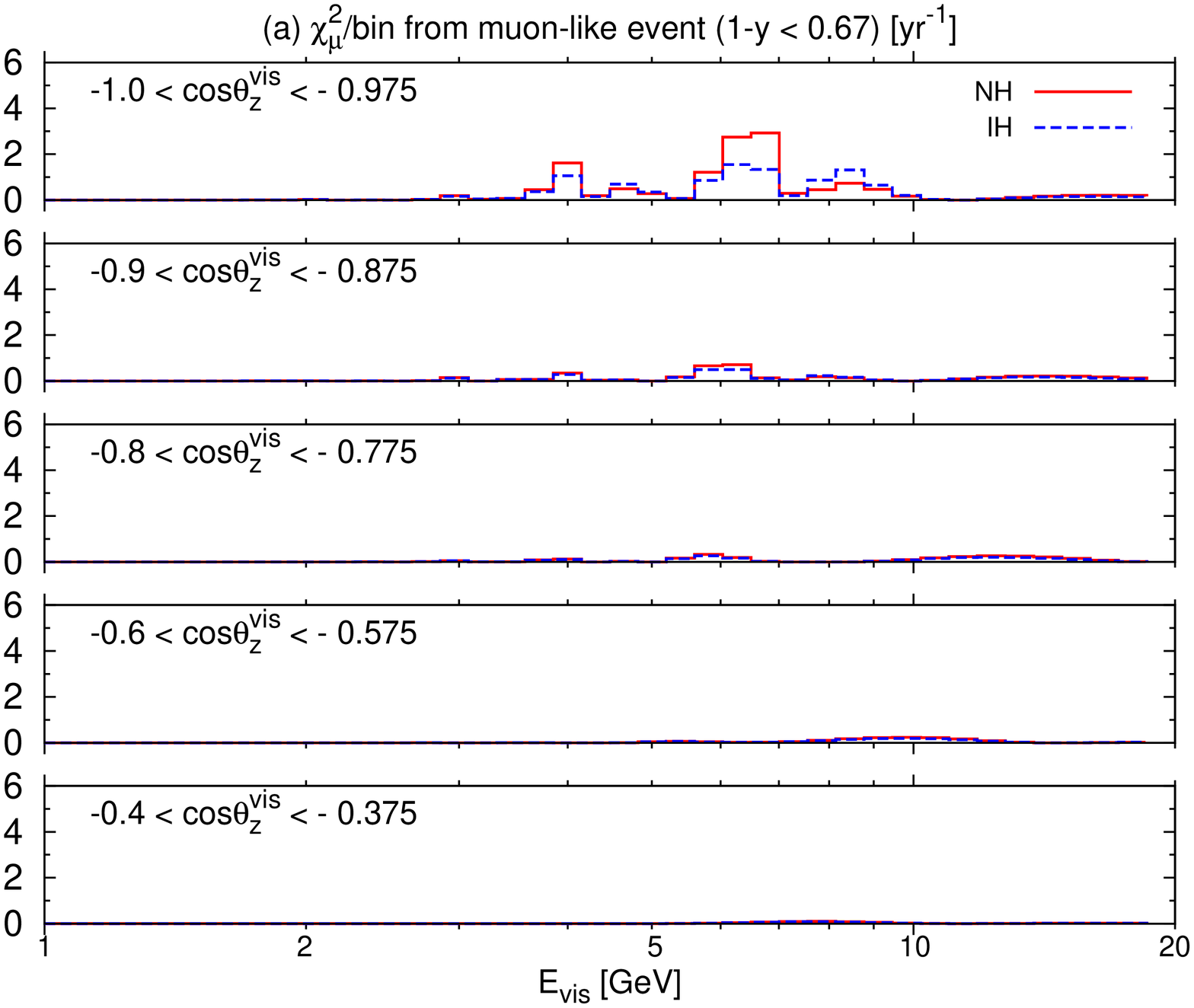}
\includegraphics[height=0.32\textwidth,width=7cm,angle=-90]{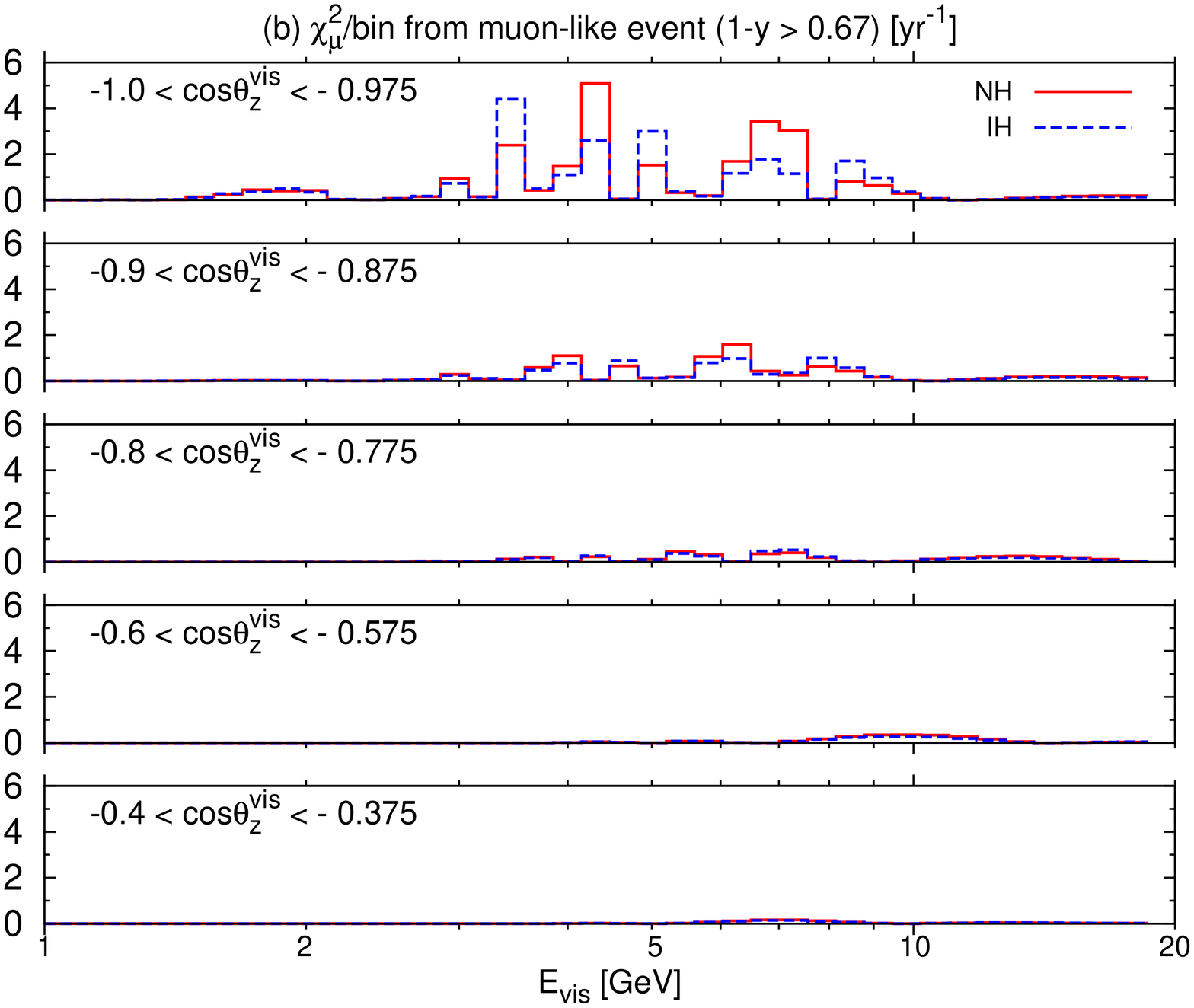}
\includegraphics[height=0.32\textwidth,width=7cm,angle=-90]{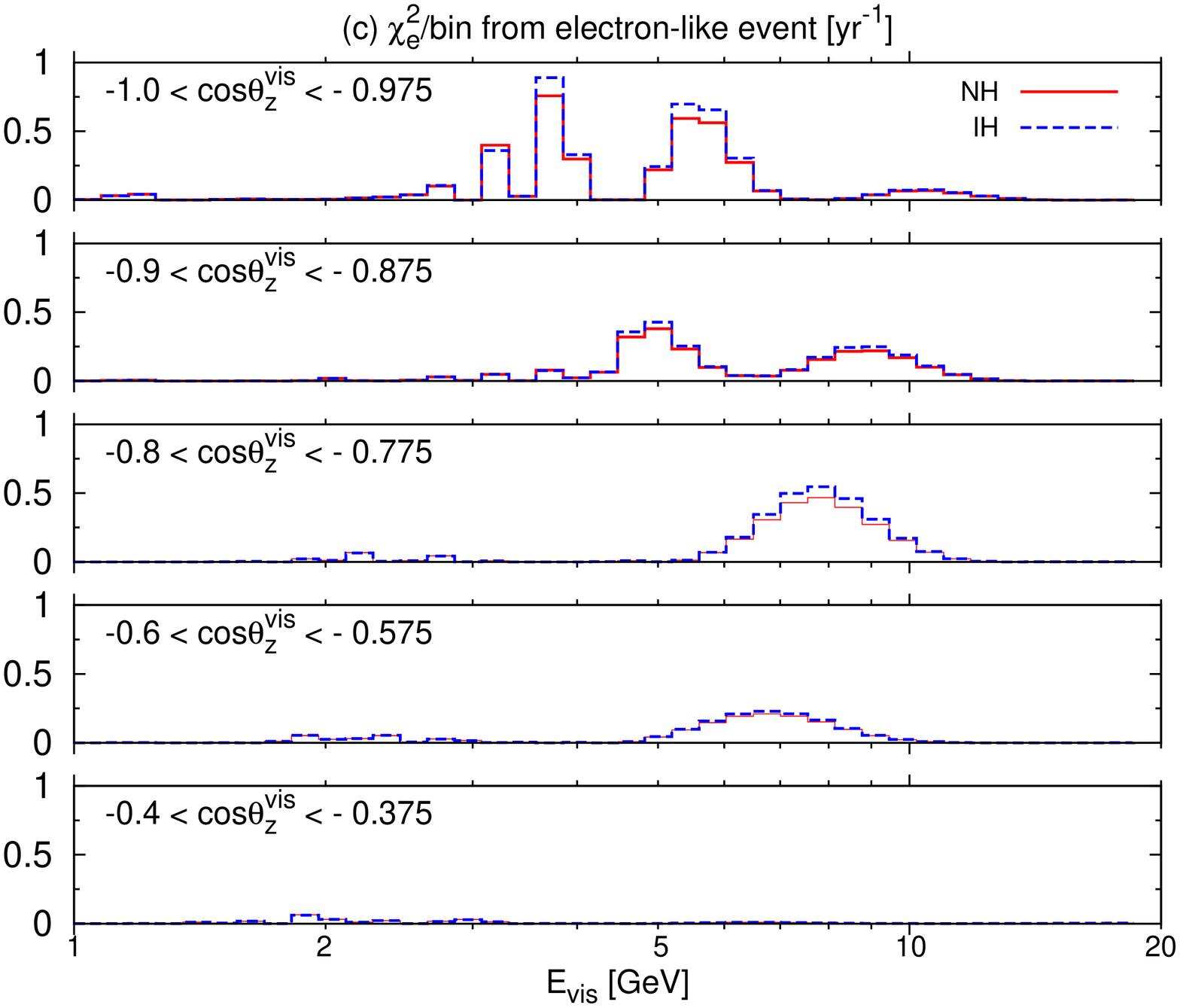}
\vspace{5mm}
\caption{Hierarchy sensitivity distribution ($\chi^2$ per bin), smeared by neutrino 
				 scattering together with energy and zenith angle
         reconstruction procedures, of (a) muon-like channel for 
         $1 - y < 0.67$, (b) muon-like channel for $1 - y > 0.67$, and (c)
         cascade channel, with NH (red-solid curves)
         and IH (blue-dashed curves), in 1-year run of PINGU.}
\label{fig:chi1}
\end{figure}

Let us first take a careful look at the muon-like events
which mainly come from $\mu$-CC while the contribution 
of $\tau$-CC is almost negligible. For the large-inelasticity muon-like
channel (a), the event rates drops down to zero for the energy range of
$E_{\rm vis} \lesssim 1.5 \, \mbox{GeV}$ with $E_\mu \lesssim 1 \, \mbox{GeV}$.
This is because of the inelasticity cut, $1 - y < 0.67$, and the muon energy cut, 
$E_\mu > 1\,\mbox{GeV}$. On the other hand, the event rates can be nonzero
in the same region for the small-inelasticity muon-like channel (b). 
As shown in (\ref{eq:Evis}), the neutrino energy can be exactly reconstructed 
for $\mu$-CC with $E_\mu > 1 \, \mbox{GeV}$ and the muon is not mis-identified. 
Nevertheless, the event rates are totally different from the muon-like neutrino 
event rates in \gfig{fig:N0}, due to scattering and the zenith angle 
reconstruction procedure which smear 
away the oscillation pattern, especially in the low-energy end, 
$E_{\rm vis} \lesssim 2 \, \mbox{GeV}$, and the horizontal region, 
$\cos \theta^{\rm vis}_{\rm z} \gtrsim - 0.5$. Of these two channels, the 
small-inelasticity muon-like channel (b) has larger event rates, from which more 
sensitivity to the neutrino mass hierarchy can be expected. 

For the cascade events, 
the largest component comes from NC which carries no information of neutrino 
oscillation and hence serves as background, while the signal comes from the 
$e$-CC and $\mu$-CC events. In $e$-CC, there are not so 
much oscillation behavior in the first place as shown in \gfig{fig:N0}. 
The $\mu$-CC contribution mainly comes from the mis-identified muon with 
$E_\mu \lesssim 1 \, \mbox{GeV}$ and hence concentrates
in the low-energy end, $E_{\rm vis} \lesssim 3 \, \mbox{GeV}$. We can expect
the $\mu$--CC contribution to the energy range, 
$E_{\rm vis} \gtrsim 3 \, \mbox{GeV}$, to increase with the muon 
mis-identification rate for $E_\mu > 1 \,, \mbox{GeV}$. Note that the 
contribution of $\tau$-CC is also negligible. 

To make the hierarchy sensitivity explicit, we show the $\chi^2$ distribution 
in \gfig{fig:chi1}. 
For all three channels, there is almost no sensitivity in the energy range of 
$E_{\rm vis} \lesssim 3 \, \mbox{GeV}$ or $E_{\rm vis} \gtrsim 10 \, \mbox{GeV}$. 
The contribution from the cascade channel can extend slightly further
into the high-energy end.
Of the two muon-like channels, the large-inelasticity one (a) 
has smaller contribution than the small-inelasticity one (b) due to lower 
statistics. Although the cascade channel
has smaller sensitivity per bin, its contribution extends from the core region
($\cos \theta^{\rm vis}_{\rm z} \lesssim - 0.84$) to 
the mantle region ($\cos \theta^{\rm vis}_{\rm z} \gtrsim - 0.84$), 
while the contribution from the muon-like channels damps quickly and can 
only span the mantle region. Consequently, the total contribution from the 
cascade channel can be comparable
with that from the muon-like channels.

\section{Energy and Angular Resolutions}
\label{sec:resolution}

Smearing effect not only comes from neutrino scattering and the 
reconstruction procedures as described in \gsec{sec:scattering}, 
but comes also from the detector resolutions of the reconstructed observables.
The actually measured value of an observable is distributed randomly around its 
true value according to the corresponding resolution function.
In this section, we analysis the basic features of energy and angular 
resolutions in \gsec{sec:Eresol} and \gsec{sec:Aresol}, respectively.
Their effects on the event rates and sensitivity distribution will be 
shown in \gsec{sec:Cresol}.

\subsection{Energy Resolution}
\label{sec:Eresol}

The Cherenkov light produced by final-state particles can only be partially 
collected by the PINGU detector. The effective area with 40-string configuration 
is around $10 m^2$ per megaton, in contrast to the huge size of a megaton 
detector which is of the characteristic scale $100 m$ with typical coverage
around 
$(4 \pi/3 \sim 6) \times (100 m)^2$ where $4 \pi/3$ corresponds to sphere and $6$ 
to cubic. A reasonable estimation of the coverage is $5 \times (100 m)^2$ per
megaton. The fraction of photons that can be collected is proportional to the
effective area and can be roughly estimated as the ratio between the effective
area and the coverage. 
In other words, about $2 \times 10^{-4}$ of the photons can be 
collected by the detection modules and most of them escape. Typically, 
$1\,\mbox{GeV}$ energy can produce approximately $1.8 \times 10^5$ Cherenkov photons. With a rate of $2 \times 10^{-4}$ detection, only $36$ photons can 
be collected. The energy fluctuation is around $\delta E / E \sim 1/6$
for $1 \, \mbox{GeV}$. To be conservative, we assume the energy resolution
to be,
\begin{equation}
  \sigma_{\rm E}
\approx
  0.2 \mbox{ GeV} \times \sqrt{E / \mbox{GeV}} \,.
\label{eq:dE}
\end{equation}
This applies to both $E_\mu$ and $E_{\rm cas}$ in (\ref{eq:Evis}),
\begin{equation}
  \mathbb P(E | E')
=
  \frac 1 {\sqrt{2 \pi} \sigma_E}
  \exp \left[ - \frac 1 2 \left( \frac {E' - E}{\sigma_E} \right)^2 \right] \,,
\end{equation}
separately. Due to this, the reconstructed 
neutrino energy for the muon-like events has a distribution around its true 
value. By comparing (\ref{eq:dE}) with the oscillation period of $N^{(0)}_\mu$
shown in \gfig{fig:N0}, we can see that the energy resolution is larger than
neutrino oscillation period in the energy range of
$E_\nu \lesssim 2 \, \mbox{GeV}$. Those pattern below $2 \, \mbox{GeV}$ can
not survive even if smearing only comes from detector resolution.

Note that, the energy resolution presented in \cite{icrc555} scales as 
$\sigma_{\rm E} \sim 0.25 \mbox{ GeV} \times (E_\nu / \mbox{GeV})$ for neutrino energy. 
This is a combination of the smearing effect due to detector resolution discussed here and the one 
from energy reconstruction procedure elaborated in \gsec{sec:Ereconst}. 
The latter has a linear scaling behavior and dominates in the high-energy end. 
Note that this only applies to the cascade channel.
Since the muon- and cascade events have different energy reconstruction
procedures, the total energy resolution is different. For muon-like 
events, only the detector resolution contributes while it is
a combination for the cascade events. Between neutrino and antineutrino
events, difference can also appear. A rigorous full-detector simulation
is needed in this regard.

\subsection{Angular Resolution}
\label{sec:Aresol}

If muon leaves a clear track in the PINGU detector, its direction can be
reconstructed with very good precision. Nevertheless, the track can be 
overwhelmed by the spherical cascade radiation if muon takes away only a small
part of the neutrino energy. We presumably take this criterion at 
$1 - y_\mu \sim 0.2$ below which the angular resolution is 50\% larger than
the one above it. The same thing also applies to those events with an 
electron. Since the radiation from electron is not so directed as the muon
track and is only slightly better than the radiation from hadronic shower,
the criterion is placed at higher energy ratio $1 - y_{\rm e} \sim 0.4$
below which the angular resolution is assigned to be 50\% larger than the one
above it. In addition, the angular resolution of electron is assigned to be 
one time larger than that of muon. 
For NC events which does not have a lepton at all and those events that lepton 
energy is too small to be recognized, the angular resolution is much worse.
As a rough approach, we use the angular distribution of CC-channel leptons 
with energy below $1 \, \mbox{GeV}$ as their angular resolution. 
These are summarized below,
\begin{eqnarray}
  \sigma_\theta
=
\begin{cases}
  1.0 \times 15^\circ \times (E_\mu / \mbox{GeV})^{-0.6} \,, & E_\mu > 1 \, \mbox{GeV}, 1 - y_\mu > 0.2 \\
  1.5 \times 15^\circ \times (E_\mu / \mbox{GeV})^{-0.6} \,, & E_\mu > 1 \, \mbox{GeV}, 1 - y_\mu < 0.2 \\
  2.0 \times 15^\circ \times (E_{\rm e} / \mbox{GeV})^{-0.6} \,, & E_{\rm e} > 1 \, \mbox{GeV}, 1 - y_{\rm e} > 0.2 \\
  3.0 \times 15^\circ \times (E_{\rm e} / \mbox{GeV})^{-0.6} \,, & E_{\rm e} > 1 \, \mbox{GeV}, 1 - y_{\rm e} < 0.2 \\
  \mathbb P(\theta^\ell_{\rm z})|_{E_\ell < 1 \, \mbox{\tiny GeV}}, & E_\ell < 1 \, \mbox{GeV}, NC, \tau-CC \mbox{ with } \tau \rightarrow \mbox{hadrons}\,.
\end{cases}
\label{eq:dT}
\end{eqnarray}

The kinematics of the smearing from angular resolution takes exactly the same 
form as the kinematics of the smearing
from neutrino scattering, shown in \gfig{fig:scattering}. The only modification
is $\vec P_\nu \rightarrow \vec P_{\rm vis}$ and 
$\vec P_{\rm vis} \rightarrow \vec P'_{\rm vis}$ where $\rm P'_{\rm vis}$ is
the actually measured visible momentum. They can be parametrized as,
\begin{equation}
- \vec P_{\rm vis}
\equiv
  \left| \vec P_{\rm vis} \right|
\left\lgroup
\begin{matrix}
  \sin \delta \theta \\
  0 \\
  \cos \delta \theta
\end{matrix}
\right\rgroup \,,
\qquad
- \vec P'_{\rm vis}
\equiv
  \left| \vec P'_{\rm vis} \right|
\left\lgroup
\begin{matrix}
  \sin \delta \theta' \cos \phi'_{\rm vis} \\
  \sin \delta \theta' \sin \phi'_{\rm vis} \\
  \cos \delta \theta'
\end{matrix}
\right\rgroup \,,
\end{equation}
in the neutrino frame where $\vec P_\nu$ aligns with the $z$-axis. 
Note that $\delta \theta$ is the opening angle between $\vec P_\nu$
and $\vec P_{\rm vis}$, determined by the neutrino scattering. Here,
the opening angle $\delta \Theta$ between $\vec P_{\rm vis}$ and
$\vec P'_{\rm vis}$ distributes according to detector resolution,
\begin{equation}
  \mathbb P(\delta \Theta)
=
  \frac{\sin \delta \Theta}{N(\sigma_\theta)}
  \exp \left[ - \frac 1 2 \left(\frac {\delta \Theta}{\sigma_\theta}\right)^2 \right] \,,
\end{equation}
where $N(\theta_\theta)$ is the normalization factor. 
For the azimuthal angle $\phi'$ of $\vec P'_{\rm vis}$ around $\vec P_{\rm vis}$, 
it is randomly distributed in the allowed range $[0, 2 \pi]$. This is a 
simplified approach since the geometry of the PINGU detector
is not isotropic and hence the angular resolution should have direction dependence.
Nevertheless, the full-detector specification is not available yet, and we adopt
this simple approximation. We find that, the main contribution
to angular smearing comes from the zenith angle reconstruction procedure 
discussed in \gsec{sec:Zrec}. Therefore, simplification in the detector angular 
resolution is not expected to introduce a significant bias.

Now the conversion formula (\ref{eq:tvis}) reads,
\begin{equation}
  \cos \delta \theta'
=
  \cos \delta \theta \cos \delta \Theta
- \sin \delta \theta \sin \delta \Theta \cos \phi' \,.
\label{eq:cTprime}
\end{equation}
For a given $\delta \theta$ between $\vec P_{\rm vis}$ and $\vec P_\nu$ as shown in
\gfig{fig:scattering}, the measured visible momentum $\vec P'_{\rm vis}$ after the
detector resolution has the $\delta \theta'$ distribution,
\begin{equation}
  \mathbb P(\delta \theta | \delta \theta')
=
  \frac 1 {2 \pi} \int 
  \frac {\mathbb P(\delta \Theta) \sin \delta \theta'}
				{\sqrt{\sin^2 \delta \theta \sin^2 \delta \Theta
        - (\cos \delta \theta \cos \delta \Theta - \cos \delta \theta')^2}} d \delta \Theta \,,
\label{eq:Pprime}
\end{equation}
around $\vec P_\nu$.

\begin{figure}[h!]
\centering
\includegraphics[height=12cm,width=7cm,angle=-90]{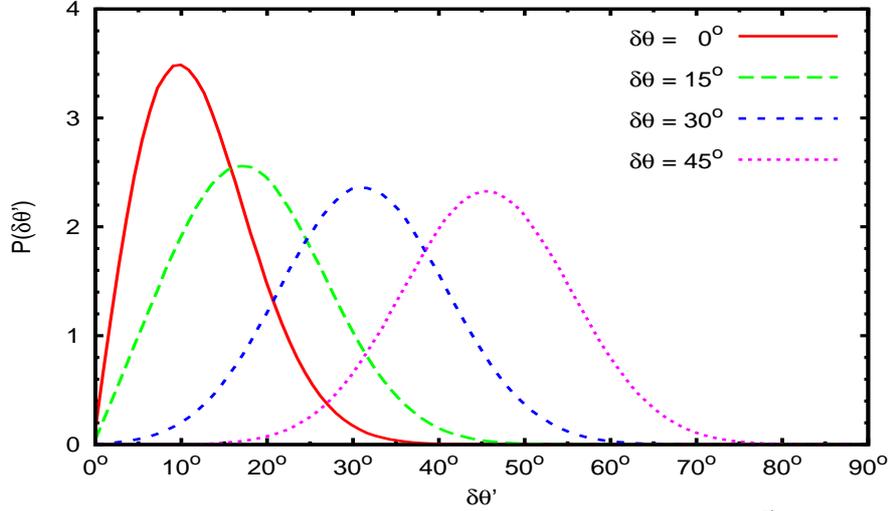}
\vspace{-4mm}
\caption{Illustration of transferring angular resolution with $\sigma_\theta = 15^\circ$ 
         from the $\vec P_{\rm vis}$ frame to the $\vec P_\nu$ frame.}
\label{fig:Aresol}
\end{figure}

\vspace{3mm}
As an illustration, we show $\mathbb P(\delta \theta | \delta \theta')$
for different values of $\delta \theta$ in \gfig{fig:Aresol}. For 
$\delta \theta = 0^\circ$, the measured angular distribution resembles the
original form, namely 
$\mathbb P(0^\circ | \delta \theta') = \mathbb P(\delta \theta')$, as indicated
by (\ref{eq:cTprime}) which becomes $\cos \delta \theta' = \cos \delta \Theta$
and hence $\delta \theta' = \delta \Theta$ under this extreme circumstance.

\subsection{Combined Effects}
\label{sec:Cresol}

\begin{figure}[h!]
\centering
\vspace{2mm}
\includegraphics[height=0.32\textwidth,width=7cm,angle=-90]{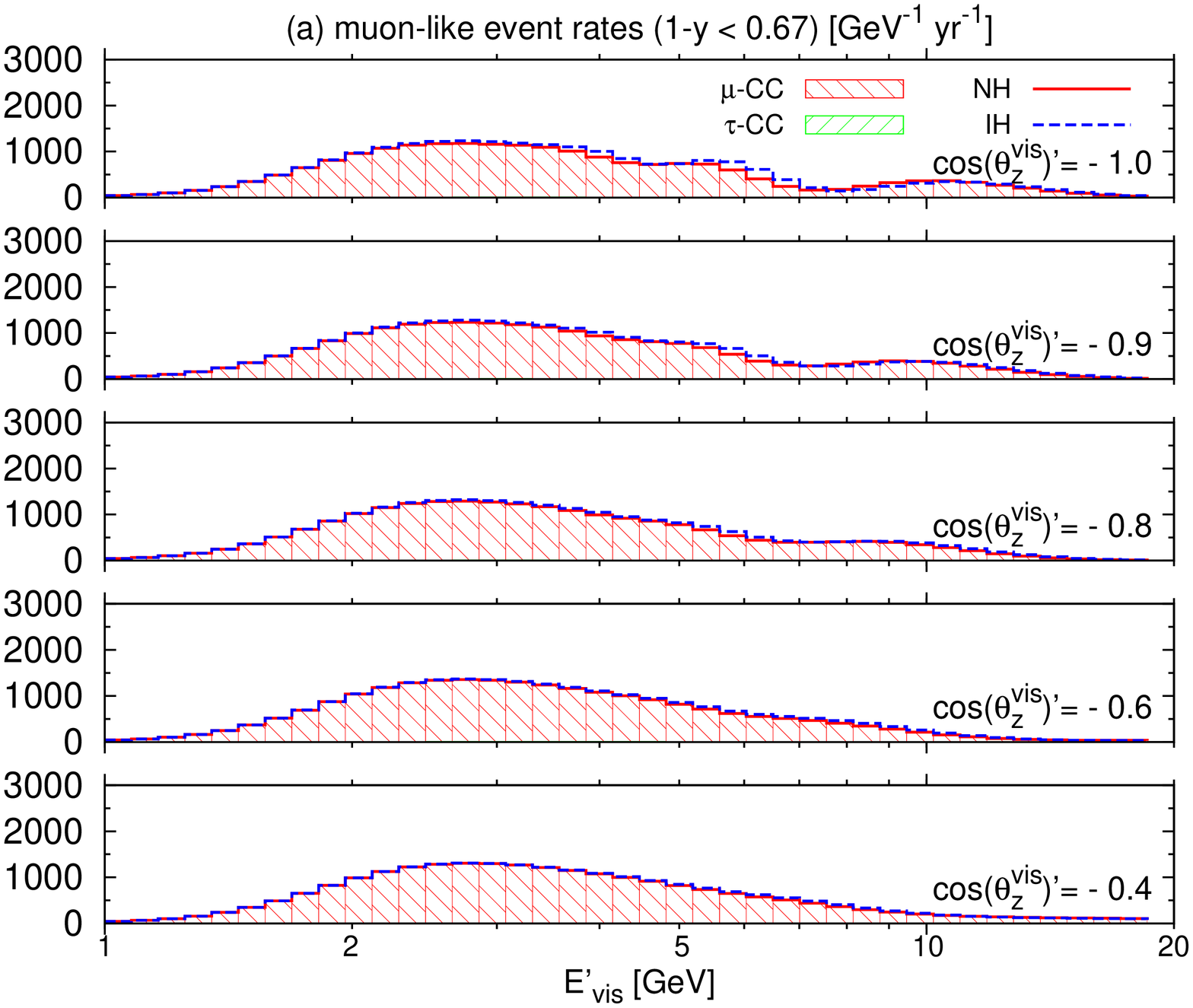}
\includegraphics[height=0.32\textwidth,width=7cm,angle=-90]{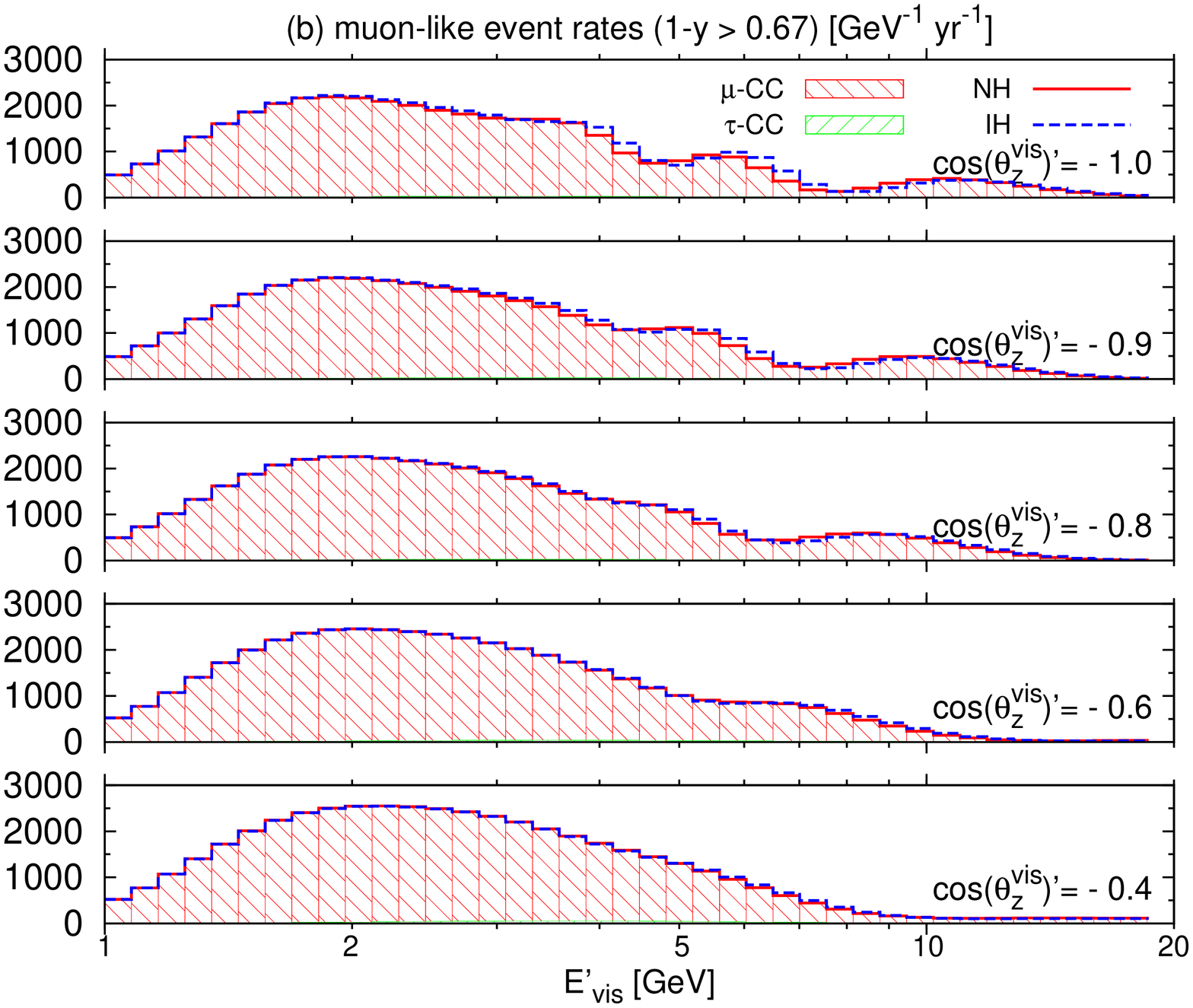}
\includegraphics[height=0.32\textwidth,width=7cm,angle=-90]{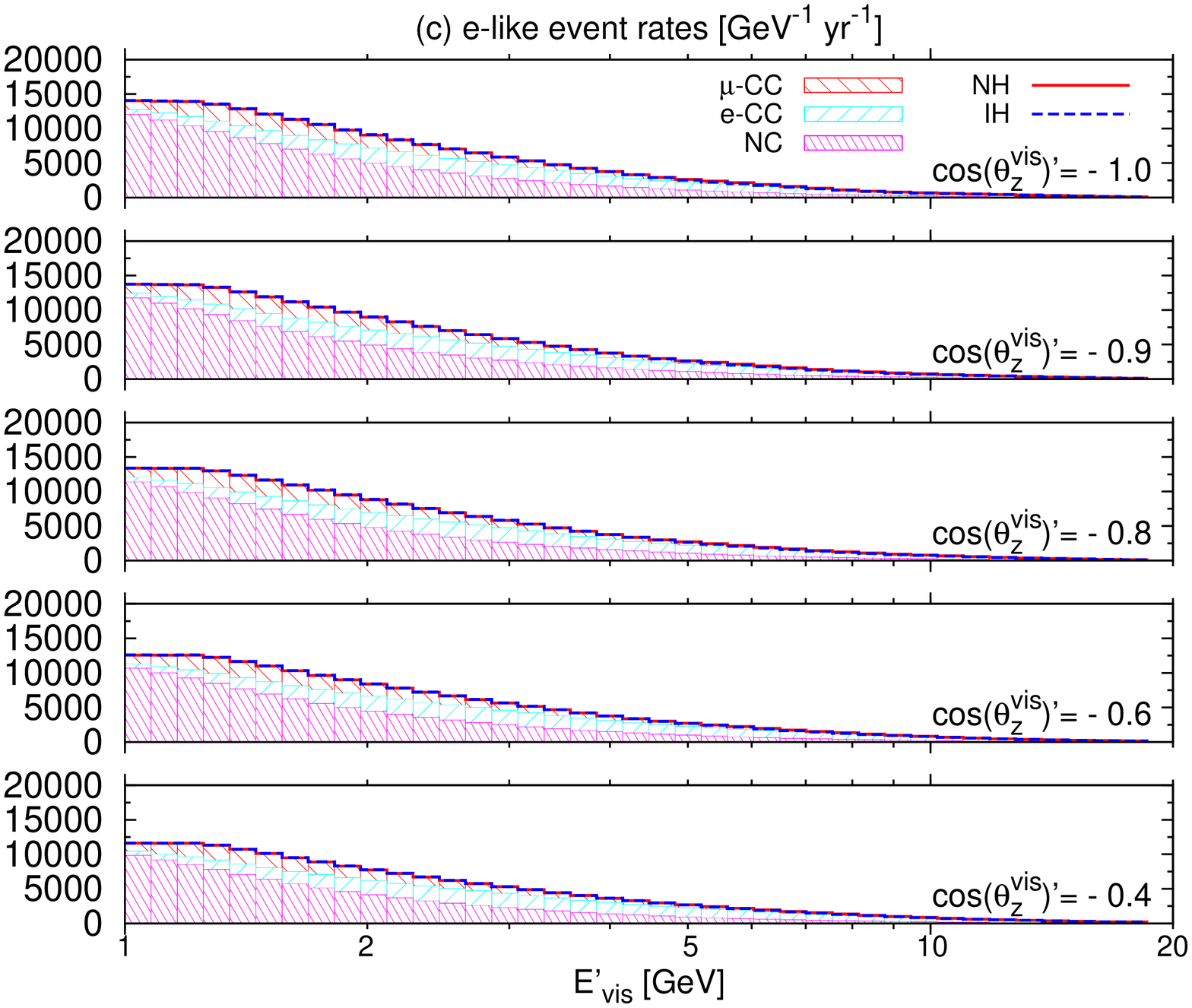}
\vspace{5mm}
\caption{Event rates, fully smeared by neutrino scattering, energy and 
				 zenith angle reconstruction procedures, as well as energy and 
         angular resolutions, of (a) muon-like channel for $1 - y < 0.67$,
         (b) muon-like channel for $1 - y > 0.67$, and (c) cascade 
         channel, with NH (red-solid curves) and IH (blue-dashed curves),
         in 1-year run of PINGU.}
\label{fig:N2}
\end{figure}

After including the effects of energy and angular resolutions, the observed 
event rates (\ref{eq:T}) become,
\begin{equation}
\hspace{-1mm}
  \frac {\partial^2 N[E'_{\rm vis}, (\theta^{\rm vis}_{\rm z})']}
				{\partial E'_{\rm vis} \partial \cos (\theta^{\rm vis}_{\rm z})'}
=
\hspace{-1mm}
  \int
  \frac {\partial^2 N(E_\nu, \theta^\nu_{\rm z})}
				{\partial E_\nu \partial \cos \theta^\nu_{\rm z}}
  \frac {\mathbb T(E_\nu| E_{\rm vis}, \delta \theta) \mathbb P(E_\mu | E'_\mu) \mathbb P(E_{\rm cas} | E'_{\rm cas}) \mathbb P(\delta \theta | \delta \theta') d E_\mu d E_{\rm cas} d \delta \theta d \delta \theta'}
				{2 \pi \sqrt{\sin^2 \theta^\nu_{\rm z} \sin^2 \delta \theta' 
        - [\cos \theta^\nu_{\rm z} \cos \delta \theta' - \cos (\theta^{\rm vis}_{\rm z})']^2}}
  d E_\nu d \cos \theta^\nu_{\rm z} \,,
\label{eq:Tprime}
\end{equation}
where $E'_{\rm vis}$ and $(\theta^{\rm vis}_{\rm z})'$ are the actually 
measured visible energy and zenith angle. Note that there are two energy 
resolution functions, one for muon and the other for cascade, since they 
can be separated. The observed event rates have been shown in \gfig{fig:N2}.
For comparison, the same scale and format as \gfig{fig:N1} are adopted.

\begin{figure}[h!]
\centering
\vspace{4mm}
\includegraphics[height=0.32\textwidth,width=7cm,angle=-90]{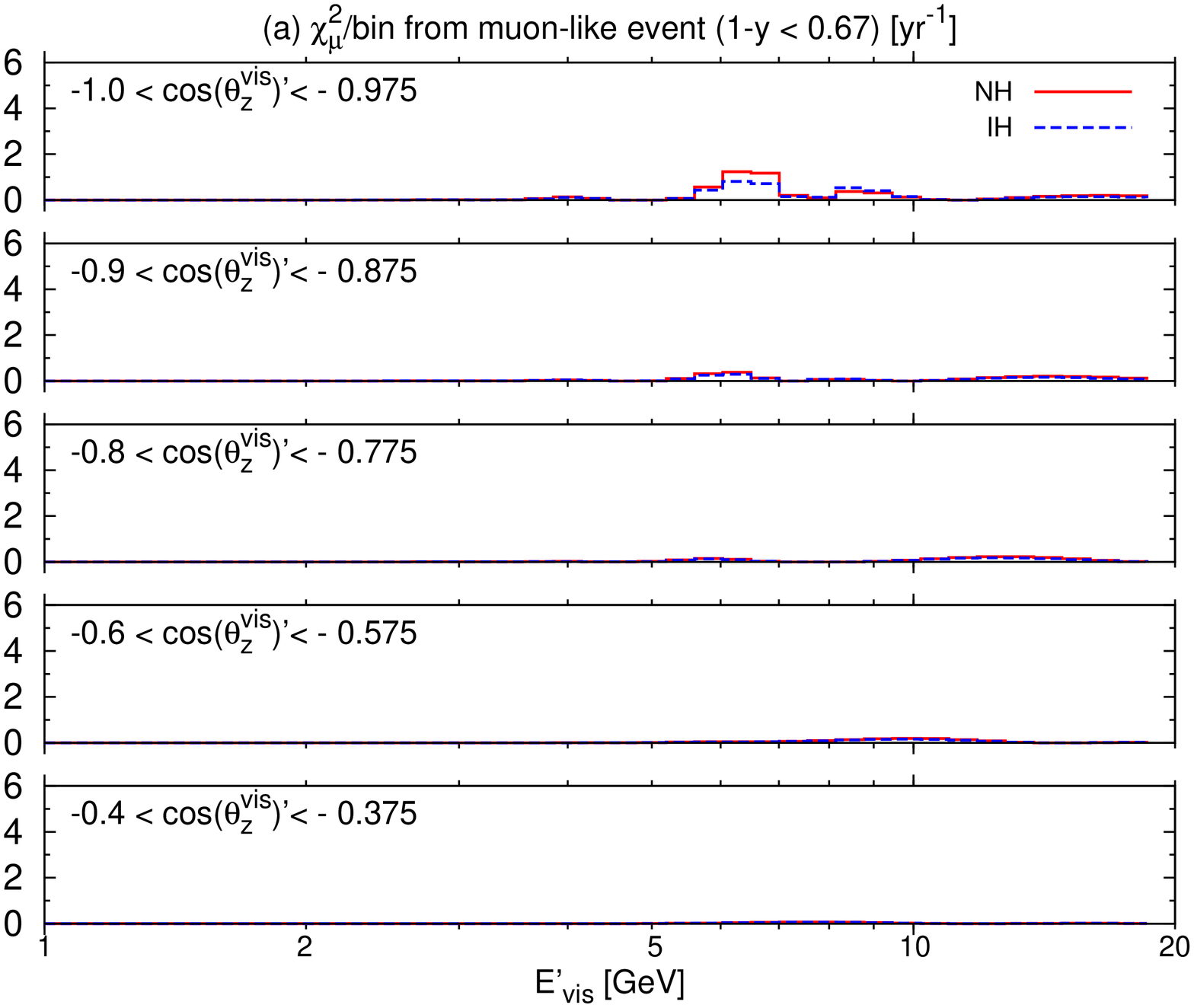}
\includegraphics[height=0.32\textwidth,width=7cm,angle=-90]{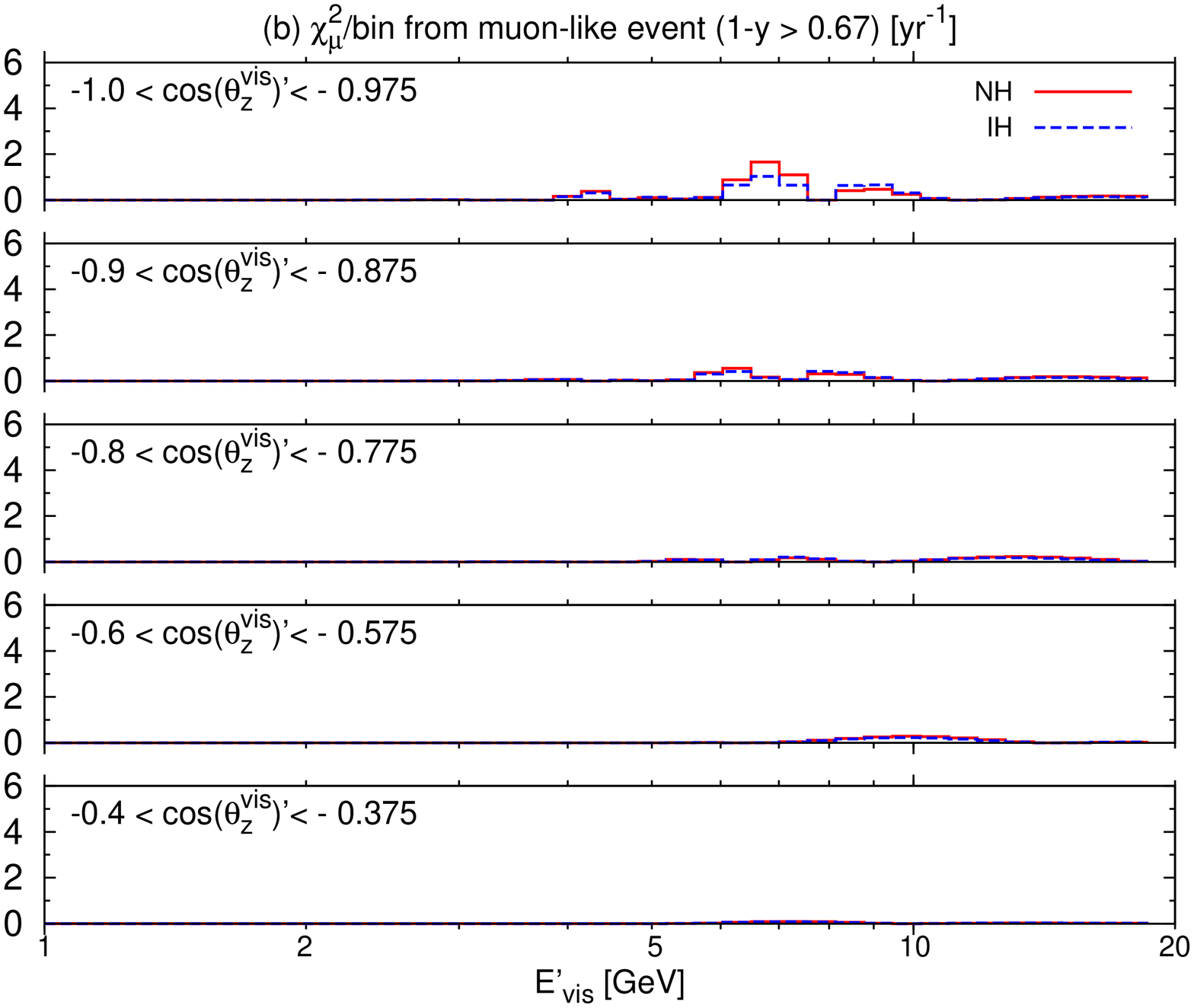}
\includegraphics[height=0.32\textwidth,width=7cm,angle=-90]{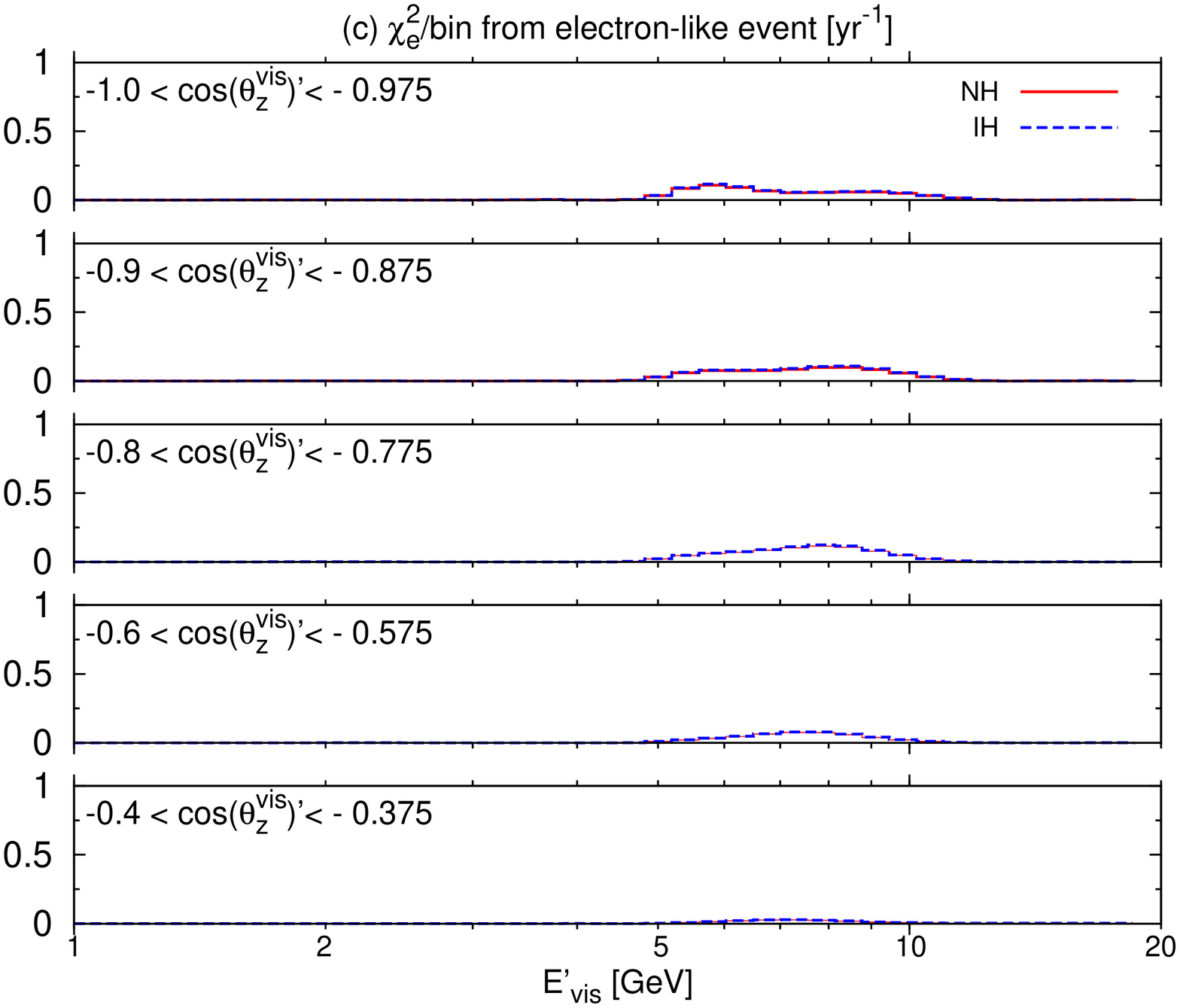}
\vspace{5mm}
\caption{Hierarchy sensitivity distribution ($\chi^2$ per bin), fully smeared by 
         neutrino scattering, energy and zenith angle reconstruction procedures,
         as well as energy and angular resolutions, of (a) muon-like channel for
         $1 - y < 0.67$, (b) muon-like channel for $1 - y > 0.67$, and (c)
         cascade channel, with NH (red-solid curves) and IH (blue-dashed
         curves), in 1-year run of PINGU.} 
\label{fig:chi2}
\end{figure}

By comparing with \gfig{fig:N1}, it needs to be noticed that the tail at the 
low-energy end extends to even lower energy due to smearing, especially for 
the muon-like events for $1-y < 0.67$. In addition, the shape becomes 
much smoother, as expected, making the sensitivity to the neutrino mass 
hierarchy vanish in the energy range of $E'_{\rm vis} \lesssim 4 \, \mbox{GeV}$. 
We can see that folding with energy and angular resolutions does not change the
event rates much, indicating that the smearing effect is dominated by neutrino
scattering and reconstruction procedures discussed in \gsec{sec:scattering}.

To make the difference clearly, the counterpart of \gfig{fig:chi1} is shown in 
\gfig{fig:chi2}. For comparison, the scale and format are kept the same. 
We can see that the sensitivity to the neutrino mass hierarchy becomes much
smaller after folding with the energy and angular resolutions. No sensitivity
in the region of $E'_{\rm vis} \lesssim 4 \, \mbox{GeV}$ or 
$\cos (\theta^{\rm vis}_{\rm z})' \gtrsim - 0.4$ survives. 
The sensitive region of the cascade events can still extends
to mantle, but the muon-like events only have sensitivity in the core.

From these observations, we can expect that 
energy and angular resolutions reduce the sensitivity to the neutrino mass 
hierarchy, but it cannot be as large as the reduction due to neutrino
scattering and reconstruction procedures.

\section{$\chi^2$ Minimization}
\label{sec:chi2}

To be consistent with our first paper \cite{Ge2013}, which is based on 
event rates at the neutrino level, we introduce the same conventional
$\chi^2$ technique,
\begin{equation}
  \chi^2
\equiv
  \sum_\alpha
  \sum_{ij} (\Delta E'_{\rm vis})_i [\Delta \cos (\theta^{\rm vis}_{\rm z})']_j
\left\{
  \dfrac { \left[ \dfrac {\partial^2 N_\alpha}{\partial E'_{\rm vis} \partial \cos (\theta^{\rm vis}_{\rm z})'} \right]^{\rm{th}}_{ij}
         - \left[ \dfrac {\partial^2 N_\alpha}{\partial E'_{\rm vis} \partial \cos (\theta^{\rm vis}_{\rm z})'} \right]^{\rm{obs}}_{ij}}
    {\sqrt{ \left[ \dfrac {\partial^2 N_\alpha}{\partial E'_\nu \partial \cos (\theta^{\rm vis}_{\rm z})'} \right]^{\rm{obs}}_{ij}}}
\right\}^2
+ \chi^2_{\rm{para}} \,,
\label{eq:chi2}
\end{equation}
to investigate the sensitivities to the neutrino mass
hierarchy, the atmospheric mixing angle $\theta_{\rm a}$ as well as its 
octant. The first term accounts for the statistics contribution from PINGU,
including three channels denoted by $\alpha$, the muon-like events for 
$1 - y \gtrless 0.67$ and the cascade events. For the visible energy,
$E'_{\rm vis}$, 40 bins are assigned logarithmically in the range from 
$1 \, \mbox{GeV}$ to $20 \, \mbox{GeV}$. The zenith angle, 
$\cos (\theta^{\rm vis}_{\rm z})'$ also has 40 bins with equal steps 
between $\cos (\theta^{\rm vis}_{\rm z})' = -1$ and $0$. Since there is not
much sensitivity for $E'_{\rm vis} \lesssim 4 \, \mbox{GeV}$ or
$\cos (\theta^{\rm vis}_{\rm z})' \gtrsim -0.4$, these regions can be cut off.
The second term, $\chi^2_{\rm para}$, includes the external constraints on
the neutrino oscillation parameters,
\begin{eqnarray}
  \chi^2_{\rm{para}}
& = &
  \left[ \frac {(\delta m^2_{\rm a})^{\rm{fit}} - \overline{\delta m^2_{\rm a}}}{\Delta \delta m^2_{\rm a}} \right]^2
+ \left[ \frac {(\delta m^2_{\rm s})^{\rm{fit}} - \overline{\delta m^2_{\rm s}}}{\Delta \delta m^2_{\rm s}} \right]^2
\nonumber
\\
& + &
  \left[ \frac {(\sin^2 2 \theta_{\rm r})^{\rm{fit}} - \overline{\sin^2 2 \theta_{\rm r}}}{\Delta \sin^2 2 \theta_{\rm r}} \right]^2
+ \left[ \frac {(\sin^2 2 \theta_{\rm s})^{\rm{fit}} - \overline{\sin^2 2 \theta_{\rm s}}}{\Delta \sin^2 2 \theta_{\rm s}} \right]^2
+ \left[ \frac {(\sin^2 2 \theta_{\rm a})^{\rm{fit}} - \overline{\sin^2 2 \theta_{\rm a}}}{\Delta \sin^2 2 \theta_{\rm a}} \right]^2 \,,
\label{eq:chi2para}
\end{eqnarray}
where the mass squared differences, $\delta m^2_{\rm a} \equiv |\delta m^2_{13}|$ and
$\delta m^2_{\rm s} \equiv \delta m^2_{21}$, the reactor mixing angle 
$\theta_{\rm r} \equiv \theta_{13}$, the solar mixing angle 
$\theta_{\rm s} \equiv \theta_{12}$, and the atmospheric mixing angle
$\theta_{\rm a} \equiv \theta_{23}$ are defined according to their physical
meanings. Their current best fit values and expected uncertainties,
\begin{subequations}
\begin{eqnarray}
&&
  \delta m^2_{\rm a} = 2.35 \pm 0.1 \times 10^{-3} \mbox{eV}^2 \,,
\qquad
  \delta m^2_{\rm s} = 7.50 \pm 0.2 \times 10^{-5} \mbox{eV}^2 \,,
\\
&&
  \sin^2 2 \theta_{\rm r} = 0.098 \pm 0.005 \,,
\qquad
  \sin^2 2 \theta_{\rm s} = 0.857 \pm 0.024 \,,
\qquad
  \sin^2 2 \theta_{\rm a} = 0.957 \pm 0.030 \,.
\end{eqnarray}
\label{eq:inputs}
\end{subequations}
\hspace{-2mm}
in the near future are taken from \cite{Machado:2011ar,minos12,pdg12} as well as
global fits \cite{Valle1205,Fogli12,Maltoni1209}. Note that some uncertainties 
are slightly smaller than the current values because improvements from
the ongoing experiments are expected before PINGU become operational.

In this section, we just consider the statistical sensitivity. 
The systematic errors will be discussed in \gsec{sec:sys}. 
The observed event rates are generated with the best fit values unless stated 
explicitly. Then the minimum of the $\chi^2$ function (\ref{eq:chi2}) can be 
obtained by varying the six neutrino
oscillation parameters, namely the two mass squared differences, the three mixing 
angles, and the CP phase, to fit the observed event rates.
Since the coefficients of $\delta$-dependent terms are small, very slight 
dependence on it can be expected. In the
following discussions, $\delta = 0^\circ$ is always adopted as its true value.
In the $\chi^2$ minimization, the parameter $\delta m^2_{\rm s}$ and 
$\theta_{\rm s}$ can not affect the result much either \cite{ino1212}.
They are fixed at their best fit values in the $\chi^2$ minimization.

\subsection{Sensitivity to the Neutrino Mass Hierarchy}

The sensitivity to the neutrino mass hierarchy can be parametrized as,
\begin{equation}
  \Delta \chi^2_{\mbox{\tiny MH}}
\equiv
  \chi^2_{\rm min} (\mbox{wrong hierarchy}) 
- \chi^2_{\rm min} (\mbox{true hierarchy})   \,,
\label{eq:chiMH}
\end{equation}
where {\it true hierarchy} is the hierarchy used to generated the observed 
event rates in (\ref{eq:chi2}) and {\it wrong hierarchy} is the opposite one.
Here we just use the Asimov data set \cite{Asimov} corresponding to 
the so-called ``{\it average experiment}'' \cite{averageExperiment}.
The statistical interpretation for such a discrete bi-value fit of mass hierarchy
can be found in \cite{stat1,stat2,stat3,stat4,stat5} which is a function of 
the $\chi^2$ function minimum, $\Delta \chi^2_{\mbox{\tiny MH}}$.

 
\begin{table}[h!]
\centering
\begin{tabular}{c||ccc|ccc}
$\Delta \chi^2_{\mbox{\tiny MH}}$ & \multicolumn{3}{c|}{NH (true)} & \multicolumn{3}{c}{IH (true)} \\
\hline
$\bar x_{\rm a}$ (true) & $- 0.2$ & $0$ & $+ 0.2$ & $- 0.2$ & $0$ & $+ 0.2$  \\
\hline\hline
\multirow{2}{*}{$\nu$} 
& \gblue{163.0} & \gblue{174.9} & \gblue{141.8} & \gblue{100.7} & \gblue{109.7} & \gblue{96.7} \\[-1mm]
&  \gred{252.9} &  \gred{215.3} &  \gred{168.9} &  \gred{143.5} &  \gred{140.7} &  \gred{120.1} 
\\
\multirow{2}{*}{Scattering \& Reconstruction} 
& \gblue{26.4} & \gblue{13.2} & \gblue{10.2} & \gblue{14.9} & \gblue{12.7} & \gblue{10.1} \\[-1mm]
&  \gred{67.3} &  \gred{32.2} &  \gred{21.3} &  \gred{21.2} &  \gred{17.9} &  \gred{15.4} 
\\
\multirow{2}{*}{Resolution} 
& \gblue{22.9} & \gblue{10.2} & \gblue{ 7.1} & \gblue{ 9.9} & \gblue{ 8.7} & \gblue{ 7.0} \\[-1mm]
&  \gred{44.1} &  \gred{17.4} &  \gred{ 9.2} &  \gred{14.8} &  \gred{13.8} &  \gred{10.0}
\\
\multirow{2}{*}{$\mu$ mis-ID} 
& \gblue{20.8} & \gblue{ 9.3} & \gblue{ 6.5} & \gblue{ 9.1} & \gblue{ 8.0} & \gblue{ 6.4} \\[-1mm]
&  \gred{40.5} &  \gred{16.0} &  \gred{ 8.5} &  \gred{14.0} &  \gred{12.9} &  \gred{ 9.3}
\\
\multirow{2}{*}{Split $\mu$ ($1-y \gtrless 0.67$)} 
& \gblue{27.1} & \gblue{12.6} & \gblue{ 8.1} & \gblue{12.8} & \gblue{10.9} & \gblue{ 7.9} \\[-1mm]
&  \gred{46.9} &  \gred{19.5} &  \gred{ 9.9} &  \gred{16.8} &  \gred{15.9} &  \gred{10.8} \\
\end{tabular}
\caption{Sensitivity to the neutrino mass hierarchy with muon-like events only 
         (blue) as well as both muon-like and cascade events (red) for 
				 different true values of the atmospheric mixing
				 angle and mass hierarchy, NH on the left and IH on the right, 
				 in 1-year run of PINGU.}
\label{tab:mh0}
\end{table}

To see the effect of each procedure discussed in previous sections, we show the results step by step. 
In the first row, the sensitivities are obtained with neutrino-level event rates, 
corresponding to our first paper \cite{Ge2013}. Then, we impose the scattering and 
reconstruction procedures elaborated in \gsec{sec:reconstruction} for
the second row. The effect of energy and angular resolutions in \gsec{sec:resolution}
can be found in the third row. Based on this, we further consider the muon mis-identification
in the fourth row. These three steps all reduce the hierarchy sensitivity to some extent.
Splitting muon-like event rates according to the inelasticity distribution in \gsec{sec:y}
can retrieve some losses as displayed in the final row.
In each case, the first sub-row in blue is 
obtained with only muon-like events and the second one in red with both muon-like
and cascade events.
Note that including the cascade events can significantly
improve the hierarchy sensitivity. 

For the whole structure, let us first take a look at the dependence on the true neutrino mass hierarchy.
If the true hierarchy is normal, the hierarchy sensitivity is higher. This trend 
starts from the result with neutrino event rates \cite{Ge2013} and can be explained 
by \gfig{fig:N0}(a) and \gfig{fig:N2}. Since the energy cut
is $E'_{\rm vis} \gtrsim 4 \, \mbox{GeV}$, the neutrino events below 
$E_\nu \sim 4 \, \mbox{GeV}$ do not contribute much. For the muon-like channel,
the event rate with IH is larger than that with NH for most of the neutrino energy 
range above $4 \, \mbox{GeV}$. With same difference in total event rates between NH 
and IH, the smaller event rates make NH more sensitive as indicated in (\ref{eq:chi2}) 
and shown in \gfig{fig:chi2}. The advantage of NH will be slightly reduced when 
cascade events are included in the analysis due to larger event rates with 
NH, see \gfig{fig:N0}(b). Although neutrino scattering, reconstruction procedures, and detector 
resolutions can smear the event rates severely, this trend is not reversed. 

For each case of NH or IH, the sensitivity depends on the true value  
$\bar x_{\rm a}$ of the atmospheric mixing angle, which has been assigned 
three values, $\pm 0.2$ and $0$. As argued in \gsec{sec:decomposition}, the 
quadratic term of $x_{\rm a}$ can dominate in the energy range 
$5 \, \mbox{GeV} \lesssim E_\nu \lesssim 10 \, \mbox{GeV}$ when 
$\bar x_{\rm a} = \pm 0.2$ and its contribution to the hierarchy sensitivity 
is destructive. This feature can modify the expected monotonic 
dependence on $\bar x_{\rm a}$ with only the linear term is considered. 
These observation are supported by the results in the first row of \gtab{tab:mh0}.
With NH, the sensitivity is larger for $\bar x_{\rm a} = - 0.2$ than for 
$\bar x_{\rm a} = 0.2$ since the quadratic term is the same but the linear term 
reduces the hierarchy sensitivity when $\bar x_{\rm a}$ increases. 
With $\bar x_{\rm a}$ switching from $0$ to $-0.2$, the 
sensitivity decreases since the quadratic term can reduce the difference between
NH and IH. This trend also applies to the case with IH.
When the cascade channel is also included, the difference between 
$\bar x_{\rm a} = -0.2$ and $\bar x_{\rm a} = +0.2$ becomes larger. Naively 
thinking, an opposition sign between the coefficients $N^{(1)}_\mu$ and 
$N^{(1)}_{\rm e}$ of the linear term of $x_{\rm a}$ can reduce the difference 
between $\bar x_{\rm a} = -0.2$ and $\bar x_{\rm a} = +0.2$ when cascade 
events are included in the analysis, in contrast to the results shown in 
\gtab{tab:mh0}. The reason is, for cascade events, the 
contribution from the linear term of $x_{\rm a}$ is relatively larger and itself 
has significant dependence on the mass hierarchy. With a negative 
$\bar x_{\rm a}$, $N^{(1)}_{\rm e}$ not only increases the total event rates, 
but more importantly enhances the difference between NH and IH. Another thing that should be 
noticed is, after including the cascade events, the monotonic dependence on 
$\bar x_{\rm a}$ appears. This is because the quadratic term only comes from 
the muon-like events. When cascade events is included, the parameter region 
in which the quadratic term dominates can be avoided in $\chi^2$ minimization.

After neutrino scattering and reconstruction procedures are
applied, the sensitivity drops significantly, by roughly an order, as expected.
The reduction in the contribution from the muon-like events is more severe
than the cascade events. This is because the muon-like event rates have
more oscillation pattern and the smearing effect is more significant than the 
cascade event rates, as shown in \gfig{fig:N0}. For the input value  
$\bar x_{\rm a}$ of the atmospheric mixing angle, monotonic dependence is restored 
even for the results with only muon-like events since after smearing the quadratic term of
$x_{\rm a}$ is no longer important across the whole energy range.

The energy and angular resolutions can further reduce the sensitivity, but the 
reduction is not so significantly. This is because the larger smearing effect comes from
the neutrino scattering and incomplete reconstruction procedures. For muon mis-identification,
the reduction is not significant either since the mis-identified muon-like events still
carry the information of neutrino mass hierarchy and contribute to the cascade events.

By splitting the muon-like events into two channels with the criterion,
$1 - y \gtrless 0.67$, the sensitivity can be significantly increased by
a factor of $13\% \sim 40\%$, consisent with \cite{Ribordy13}. 
It can compensate the sensitivity reduction
due to detector resolutions and muon mis-identification for muon-like events.

\subsection{Precision on the Atmospheric Mixing Angle}

To obtain the precision on the atmospheric mixing angle $\theta_{\rm a}$, we 
replace its external constraint, namely the last term in (\ref{eq:chi2para}), by
\begin{equation}
\left[
  \frac { \left( \sin^2 2 \theta_{\rm a} \right)^{\rm fit}
        - \overline{\sin^2 2 \theta_{\rm a}} }
				{ \Delta \sin^2 2 \theta_{\rm a} }
\right]^2
\rightarrow
\left[ \frac {x^2_{\rm a} - \bar x^2_{\rm a}}{0.03} \right]^2 \,,
\end{equation}
in order to avoid artificial contribution from the assumed true value of 
the atmospheric mixing angle. In other words, we keep the $1 \, \sigma$ 
error of the present constraint in (\ref{eq:inputs}) around the input values 
instead of the best fit value.
For each true value $\bar x_{\rm a}$, the $\chi^2$ minimization is carried out by fixing 
the fitting parameter $x_{\rm a}$. The resultant $\chi^2_{\rm min}(x_{\rm a})$ 
is hence a function of the fitting parameter $x_{\rm a}$, from which the 
precision on $x_{\rm a}$ can be determined \cite{Ge2013}.
In \gtab{tab:xa0} we shown the results.

\begin{table}[h!]
\centering
\begin{tabular}{c||ccc|ccc}
$\Delta(x_{\rm a})$ & \multicolumn{3}{c|}{NH (true)} & \multicolumn{3}{c}{IH (true)} \\
\hline
$\bar x_{\rm a}$ (true) & $- 0.2$ & $0$ & $+ 0.2$ & $- 0.2$ & $0$ & $+ 0.2$  \\
\hline\hline
\multirow{2}{*}{$\nu$} 
& \gblue{0.014} & \gblue{0.036} & \gblue{0.011} & \gblue{0.014} & \gblue{0.046} & \gblue{0.011} \\[-1mm]
&  \gred{0.012} &  \gred{0.025} &  \gred{0.010} &  \gred{0.012} &  \gred{0.035} &  \gred{0.011} 
\\
\multirow{2}{*}{Scattering \& Reconstruction} 
& \gblue{0.025} & \gblue{0.051} & \gblue{0.015} & \gblue{0.024} & \gblue{0.073} & \gblue{0.017} \\[-1mm]
&  \gred{0.020} &  \gred{0.039} &  \gred{0.014} &  \gred{0.022} &  \gred{0.061} &  \gred{0.016} 
\\
\multirow{2}{*}{Resolution} 
& \gblue{0.027} & \gblue{0.055} & \gblue{0.016} & \gblue{0.025} & \gblue{0.077} & \gblue{0.018} \\[-1mm]
&  \gred{0.021} &  \gred{0.043} &  \gred{0.016} &  \gred{0.024} &  \gred{0.067} &  \gred{0.018} 
\\
\multirow{2}{*}{$\mu$ mis-ID} 
& \gblue{0.028} & \gblue{0.059} & \gblue{0.017} & \gblue{0.026} & \gblue{0.078} & \gblue{0.019} \\[-1mm]
&  \gred{0.022} &  \gred{0.045} &  \gred{0.017} &  \gred{0.025} &  \gred{0.070} &  \gred{0.019}
\\
\multirow{2}{*}{Split $\mu$ ($1-y \gtrless 0.67$)} 
& \gblue{0.027} & \gblue{0.057} & \gblue{0.017} & \gblue{0.026} & \gblue{0.077} & \gblue{0.019} \\[-1mm]
&  \gred{0.022} &  \gred{0.045} &  \gred{0.016} &  \gred{0.024} &  \gred{0.068} &  \gred{0.019} \\ 
\end{tabular}
\caption{Precision on the atmospheric mixing angle for muon-like events only 
         (blue) as well as both muon-like and cascade events (red) for 
				 different true values of the atmospheric mixing
				 angle and mass hierarchy, NH on the left and IH on the right, 
				 in 1-year run of PINGU.}
\label{tab:xa0}
\end{table}

As demonstrated in \cite{Ge2013}, the precision $\Delta(x_{\rm a})$ is mainly 
dictated by the combined coefficients, 
$N^{(1)}_\mu + 2 \bar x_{\rm a} N^{(5)}_\mu$, where $\bar x_{\rm a}$ is the 
input (true) value of the atmospheric mixing angle. Since $N^{(5)}_\mu$ is much
larger than $N^{(1)}_\mu$ as shown in \gfig{fig:N0}, the second term dominates
for $\bar x_{\rm a} = \pm 0.2$, leading to comparable precisions for 
$\bar x_{\rm a} = -0.2$ and $\bar x_{\rm a} = + 0.2$ with small difference due 
to the first term $N^{(1)}_\mu$. For all cases, the precision with vanishing
$\bar x_{\rm a}$ is the largest since the second term 
$2 \bar x_{\rm a} N^{(5)}_\mu$ vanishes. 
This pattern remains even after the cascade events are included, 
but slightly reduced. Including cascade events can help to enhance
the precision on $x_{\rm a}$, especially when it is small, 
$\bar x_{\rm a} \approx 0$.

Of the three steps applied to the neutrino event rates, neutrino scattering and 
reconstruction procedures have the largest impact on reducing the precision on $x_{\rm a}$.
It is quite stable when muon-like events are split into two parts, and when 
detector resolutions and muon mis-identification are imposed.

\subsection{Sensitivity to the Octant of the Atmospheric Mixing Angle}

\begin{table}[h!]
\centering
\begin{tabular}{c||cccc|cccc}
$\Delta \chi^2_{\mbox{\tiny Octant}}$ & \multicolumn{4}{c|}{NH (true)} & \multicolumn{4}{c}{IH (true)} \\
\hline
$\bar x_{\rm a}$ (true) & $- 0.2$ & $-0.1$ & $+0.1$ & $+ 0.2$ & $- 0.2$ & $-0.1$ & $+0.1$ & $+ 0.2$  \\
\hline\hline
\multirow{2}{*}{$\nu$} 
& \gblue{41.5} & \gblue{6.7} & \gblue{20.6} & \gblue{64.1} & \gblue{10.3} & \gblue{2.9} & \gblue{4.1} & \gblue{12.9} \\[-1mm]
&  \gred{84.8} & \gred{15.1} &  \gred{29.3} & \gred{162.6} &  \gred{32.5} &  \gred{8.0} & \gred{12.0} &  \gred{46.0}
\\
\multirow{2}{*}{Scattering \& Reconstruction} 
& \gblue{24.4} & \gblue{3.1} & \gblue{9.6} & \gblue{47.7} & \gblue{7.5} & \gblue{1.8} & \gblue{2.8} & \gblue{9.9} \\[-1mm]
&  \gred{36.9} &  \gred{6.2} & \gred{13.2} & \gred{103.0} & \gred{15.6} &  \gred{3.3} &  \gred{5.5} & \gred{21.4}
\\
\multirow{2}{*}{Resolution} 
& \gblue{20.4} & \gblue{2.6} & \gblue{8.3} & \gblue{43.4} & \gblue{6.3} & \gblue{1.4} & \gblue{2.5} & \gblue{8.8} \\[-1mm]
&  \gred{27.9} &  \gred{4.4} & \gred{10.2} &  \gred{76.2} & \gred{11.1} &  \gred{2.3} &  \gred{4.1} & \gred{15.4}
\\
\multirow{2}{*}{$\mu$ mis-ID} 
& \gblue{18.6} & \gblue{2.4} & \gblue{7.5} & \gblue{39.2} & \gblue{5.7} & \gblue{1.3} & \gblue{2.2} & \gblue{7.9} \\[-1mm]
&  \gred{25.9} &  \gred{4.1} &  \gred{9.2} &  \gred{69.9} & \gred{10.3} &  \gred{2.1} &  \gred{3.7} & \gred{14.3}
\\
\multirow{2}{*}{Split $\mu$ ($1-y \gtrless 0.67$)} 
& \gblue{19.6} & \gblue{2.5} & \gblue{7.8} & \gblue{40.4} & \gblue{6.6} & \gblue{1.5} & \gblue{2.5} & \gblue{8.7} \\[-1mm]
&  \gred{26.9} &  \gred{4.2} &  \gred{9.5} &  \gred{70.9} & \gred{11.2} &  \gred{2.3} &  \gred{4.0} & \gred{15.1} \\
\end{tabular}
\caption{Sensitivity to the octant of the atmospheric mixing angle for muon-like 
				 events only 
         (blue) as well as both muon-like and cascade events (red) for 
				 different true values of the atmospheric mixing
				 angle and mass hierarchy, NH on the left and IH on the right, 
				 in 1-year run of PINGU.}
\label{tab:oct0}
\end{table}

The octant sensitivity can be defined just like the sensitivity to the neutrino
mass hierarchy (\ref{eq:chiMH}),
\begin{equation}
  \Delta \chi^2_{\mbox{\tiny Octant}}
\equiv
  |\chi^2_{\rm min} (\mbox{LO}) - \chi^2_{\rm min} (\mbox{HO})| \,,
\end{equation}
where $\chi^2_{\rm min} (\mbox{LO})$ is obtained by limiting the fit parameter
with $\theta_{\rm a} < 45^\circ$ and $\chi^2_{\rm min} (\mbox{HO})$ 
with $\theta_{\rm a} > 45^\circ$. The results are shown in \gtab{tab:oct0}
for four true values $\bar x_{\rm a} = \pm 0.2, \, \pm 0.1$.
We can see that the octant sensitivity is larger if the true hierarchy is normal, 
due to smaller total event rates and also larger linear term coefficients 
$N^{(1)}_\alpha$ ($\alpha = \mu, {\rm e}$) with NH, and if the atmospheric 
angle is in the lower octant with a positive $\bar x_{\rm a}$. The sensitivity 
can be effectively reduced due to neutrino scattering and reconstruction 
procedures, as well as resolutions and muon mis-identification, and slightly 
increased after splitting the muon-like events into two channels according to the 
inelasticity distribution. Note that the cascade events can significantly 
improve the octant sensitivity.

\section{Systematic Errors}
\label{sec:sys}


In this section, we study the impacts of systematic errors in the energy 
and angular resolutions, the muon mis-identification rate, and the overall 
normalization of the flux times cross sections, to examine how each of them 
affects the measurement sensitivities. The impacts of combining these four 
systematic errors have also been studied. Note that this list of systematic 
uncertainties is far from complete. For instance, there can be uncertainties 
in neutrino/antineutrino ratios (fluxes, cross sections), event energy shape 
and scale, and so on. A complete 
survey of these systematic uncertainties is beyond the scope of this paper.

The details of the systematics we consider are listed below.
\begin{enumerate}
\item For the energy 
resolution, we keep the Gaussian form (\ref{eq:dE}) and assign a 15\% error, namely 
$\sigma_{\mbox{\tiny E}} = (0.2 \pm 0.03) \mbox{ GeV} \times \sqrt{E/\mbox{GeV}}$.

\item For the angular resolution, any parameters in the functional form (\ref{eq:dT}) 
can suffer from systematic uncertainties. According to the physical picture illustrated 
in \gsec{sec:Aresol}, we just examine the impacts of allowing 20\% uncertainty in the
common resolution of $\sigma_\theta$, i.e.,
$\sigma_\theta = (15^\circ \pm 3^\circ) \times (E/\mbox{GeV})^{-0.6}$ for 
the first case in (\ref{eq:dT}) while retaining the relation between different cases.

\item The muon mis-identification rate should naively be small at high muon energies.
Since full-detector simulation of the PINGU detector is not available to us,
we simply consider a constant 20\% error in the constant mis-identification rate, i.e.,
$(10 \pm 2) \%$ mis-identification rate which is independent of the muon energy.

\item
For the overall normalization, we assign a common 5\% error for all the neutrino 
events, since we expect that the largest error comes from the atmospheric neutrino 
fluxes. Although muon- and electron-neutrino as well as anti-neutrino fluxes may each have 
distinct uncertainties, we expect that the relative errors to be small and the common
overall normalization error dominates \cite{Stanev04} because they have common origins. 
Experimentally, the normalization uncertainty can be reduced by measuring both up- and 
down-going neutrinos
at PINGU and using down-going event rates to normalize the up-going one, or by 
anchoring the flux at high energies ($E_\nu > 20 \, \mbox{GeV}$) \cite{Franco13}.
\end{enumerate}

We use the so called {\it pull method} \cite{pullMethod} to treat the systematic errors.
Their impacts on the event rates can be parametrized as linear functions,
\begin{equation}
  \frac {\partial^2 N_\alpha (c_i)}{\partial E'_{\rm vis} \partial \cos (\theta^{\rm vis}_{\rm z})'}
\approx
  \frac {\partial^2 N_\alpha \left( c^{(0)}_i \right)}{\partial E'_{\rm vis} \partial \cos (\theta^{\rm vis}_{\rm z})'}
+ 
  \sum_i
  \partial_{c_i} 
\left. \left[ \frac {\partial^2 N_\alpha (c_i)}{\partial E'_{\rm vis} \partial \cos (\theta^{\rm vis}_{\rm z})'} \right] \right|_{c_i = c^{(0)}_i} 
  \times \left( c_i - c^{(0)}_i \right) \,,
\end{equation}
where $c_i$ parametrizes the variable with a systematic error,
and $c^{(0)}_i$ represents the corresponding central value. This linearized 
approximation can hold quite well for all the four systematic errors considered
here.
Their effects, when imposed separately or combined, on the sensitivity to mass hierarchy, 
the precision on the atmospheric mixing angle, and the sensitivity to the octant, 
have been summarized in \gtab{tab:mh1}, \gtab{tab:xa1}, and \gtab{tab:oct1}, 
respectively.

\begin{table}[h!]
\centering
\begin{tabular}{c||ccc|ccc}
$\Delta \chi^2_{\mbox{\tiny MH}}$ & \multicolumn{3}{c|}{NH (true)} & \multicolumn{3}{c}{IH (true)} \\
\hline
$\bar x_{\rm a}$ (true) & $- 0.2$ & $0$ & $+ 0.2$ & $- 0.2$ & $0$ & $+ 0.2$  \\
\hline\hline
\multirow{2}{*}{$\sigma_{\mbox{\tiny E}} = (0.2 \pm 0.03) \mbox{ GeV} \times \sqrt{E/\mbox{GeV}}$} 
& \gblue{26.1} & \gblue{11.2} & \gblue{7.7} & \gblue{11.5} & \gblue{10.0} & \gblue{7.1} \\[-1mm]
&  \gred{42.3} &  \gred{18.0} &  \gred{8.6} &  \gred{16.0} &  \gred{14.8} &  \gred{9.9} 
\\
\multirow{2}{*}{$\sigma_\theta = (15^\circ \pm 3^\circ) \times (E/\mbox{GeV})^{-0.6}$} 
& \gblue{26.3} & \gblue{11.5} & \gblue{7.9} & \gblue{11.8} & \gblue{ 9.9} & \gblue{ 7.6} \\[-1mm]
&  \gred{43.1} &  \gred{18.2} &  \gred{9.0} &  \gred{16.1} &  \gred{15.1} &  \gred{10.0} 
\\
\multirow{2}{*}{$\mu$ mis-ID ($10\% \pm 2\%$)} 
& \gblue{19.0} & \gblue{ 8.4} & \gblue{6.1} & \gblue{ 8.8} & \gblue{ 7.2} & \gblue{5.3} \\[-1mm]
&  \gred{42.1} &  \gred{17.1} &  \gred{8.5} &  \gred{15.3} &  \gred{13.3} &  \gred{9.2}
\\
\multirow{2}{*}{{\small Normalization} ($1 \pm 0.05$)} 
& \gblue{16.8} & \gblue{ 7.9} & \gblue{6.6} & \gblue{ 8.6} & \gblue{ 6.9} & \gblue{5.0} \\[-1mm]
&  \gred{43.6} &  \gred{16.9} &  \gred{8.7} &  \gred{14.0} &  \gred{13.8} &  \gred{8.7} 
\\
\hline
\multirow{2}{*}{Combined} 
& \gblue{12.6} & \gblue{ 7.5} & \gblue{5.9} & \gblue{ 7.5} & \gblue{ 6.2} & \gblue{4.8} \\[-1mm]
&  \gred{41.6} &  \gred{16.1} &  \gred{8.3} &  \gred{13.7} &  \gred{12.2} &  \gred{7.7} 
\end{tabular}
\caption{The impact of systematic errors on the sensitivity to the neutrino mass 
				 hierarchy with only muon-like events (blue) as well as both muon-like
				 and cascade events (red) for 
				 different true values of the atmospheric mixing
				 angle and mass hierarchy, NH on the left and IH on the right, 
				 in 1-year run of PINGU.}
\label{tab:mh1}
\end{table}


By comparing \gtab{tab:mh1} with \gtab{tab:mh0}, we observe that the results 
are quite stable under the systematic error in energy and angular resolutions
as expected from the fact that the smearing effect mainly comes from the neutrino
scattering together with the reconstruction procedures, and the results are 
not much affected by the detector resolutions in the first place. 
For the systematic error in the muon mis-identification rate, if only 
the muon-like events are used (blue), the sensitivity is significantly reduced 
by the 20\% uncertainty. The situation becomes stable when the cascade 
events are also included in the analysis (red), because
the mis-identified muon still carries the information of atmospheric neutrino 
oscillation which is not lost but contributes to the cascade events 
instead. The error from the overall normalization has the largest influence for 
muon-like events because the lower or higher flux can mimic the presence or 
absence of the MSW resonance in the dominant muon channel. The negative impacts 
can be partially recovered by including the cascade events.

\begin{table}[h!]
\centering
\begin{tabular}{c||ccc|ccc}
$\Delta(x_{\rm a})$ & \multicolumn{3}{c|}{NH (true)} & \multicolumn{3}{c}{IH (true)} \\
\hline
$\bar x_{\rm a}$ (true) & $- 0.2$ & $0$ & $+ 0.2$ & $- 0.2$ & $0$ & $+ 0.2$  \\
\hline\hline
\multirow{2}{*}{$\sigma_{\mbox{\tiny E}} = (0.2 \pm 0.03) \mbox{ GeV} \times \sqrt{E/\mbox{GeV}}$} 
& \gblue{0.027} & \gblue{0.057} & \gblue{0.017} & \gblue{0.026} & \gblue{0.077} & \gblue{0.019} \\[-1mm]
&  \gred{0.025} &  \gred{0.045} &  \gred{0.016} &  \gred{0.025} &  \gred{0.068} &  \gred{0.019} 
\\
\multirow{2}{*}{$\sigma_\theta = (15^\circ \pm 3^\circ) \times (E/\mbox{GeV})^{-0.6}$} 
& \gblue{0.028} & \gblue{0.057} & \gblue{0.017} & \gblue{0.026} & \gblue{0.078} & \gblue{0.019} \\[-1mm]
&  \gred{0.023} &  \gred{0.045} &  \gred{0.017} &  \gred{0.024} &  \gred{0.069} &  \gred{0.019} 
\\
\multirow{2}{*}{$\mu$ mis-ID ($10\% \pm 2\%$)} 
& \gblue{0.028} & \gblue{0.069} & \gblue{0.021} & \gblue{0.028} & \gblue{0.083} & \gblue{0.024} \\[-1mm]
&  \gred{0.022} &  \gred{0.052} &  \gred{0.020} &  \gred{0.026} &  \gred{0.078} &  \gred{0.020} 
\\
\multirow{2}{*}{{\small Normalization} ($1 \pm 0.05$)} 
& \gblue{0.033} & \gblue{0.081} & \gblue{0.025} & \gblue{0.031} & \gblue{0.090} & \gblue{0.030} \\[-1mm]
&  \gred{0.023} &  \gred{0.055} &  \gred{0.022} &  \gred{0.027} &  \gred{0.078} &  \gred{0.021} 
\\
\hline
\multirow{2}{*}{Combined} 
& \gblue{0.034} & \gblue{0.083} & \gblue{0.028} & \gblue{0.033} & \gblue{0.099} & \gblue{0.032} \\[-1mm]
&  \gred{0.026} &  \gred{0.058} &  \gred{0.024} &  \gred{0.029} &  \gred{0.081} &  \gred{0.023} 
\end{tabular}
\caption{The impact of systematic errors on the precision of the atmospheric 
				 mixing angle $\theta_{\rm a}$ with only muon-like events (blue) as well 
				 as both muon-like and cascade events (red) for 
				 different true values of the atmospheric mixing
				 angle and mass hierarchy, NH on the left and IH on the right, 
				 in 1-year run of PINGU.}
\label{tab:xa1}
\end{table}


On the other hand, comparison of \gtab{tab:xa0} and \gtab{tab:xa1} shows that 
the precision on the atmospheric mixing angle parameter $x_{\rm a} = 1/2 - \sin^2 \theta_{\rm a}$
is very stable under the influence of all the systematic errors, with the only 
exception of the systematic error in the overall normalization when only the muon-like
events are considered.

\begin{table}[h!]
\centering
\begin{tabular}{c||cccc|cccc}
$\Delta \chi^2_{\mbox{\tiny Octant}}$ & \multicolumn{4}{c|}{NH (true)} & \multicolumn{4}{c}{IH (true)} \\
\hline
$\bar x_{\rm a}$ (true) & $- 0.2$ & $-0.1$ & $+0.1$ & $+ 0.2$ & $- 0.2$ & $-0.1$ & $+0.1$ & $+ 0.2$  \\
\hline\hline
\multirow{2}{*}{$\sigma_{\mbox{\tiny E}} = (0.2 \pm 0.03) \mbox{ GeV} \times \sqrt{E/\mbox{GeV}}$} 
& \gblue{19.4} & \gblue{2.5} & \gblue{7.7} & \gblue{40.0} & \gblue{6.5} & \gblue{1.5} & \gblue{2.5} & \gblue{8.7} \\[-1mm]
&  \gred{26.3} &  \gred{4.1} &  \gred{9.1} &  \gred{65.8} & \gred{10.8} &  \gred{2.3} &  \gred{3.8} & \gred{14.3}
\\
\multirow{2}{*}{$\sigma_\theta = (15^\circ \pm 3^\circ) \times (E/\mbox{GeV})^{-0.6}$} 
& \gblue{19.3} & \gblue{2.5} & \gblue{7.7} & \gblue{40.2} & \gblue{6.6} & \gblue{1.5} & \gblue{2.5} & \gblue{8.7} \\[-1mm]
&  \gred{26.7} &  \gred{4.2} &  \gred{9.1} &  \gred{66.8} & \gred{11.0} &  \gred{2.3} &  \gred{3.8} & \gred{14.5}
\\
\multirow{2}{*}{$\mu$ mis-ID ($10\% \pm 2\%$)} 
& \gblue{19.1} & \gblue{2.2} & \gblue{5.2} & \gblue{39.2} & \gblue{4.5} & \gblue{1.3} & \gblue{1.9} & \gblue{8.3} \\[-1mm]
&  \gred{26.5} &  \gred{4.0} &  \gred{8.2} &  \gred{66.1} & \gred{10.8} &  \gred{2.0} &  \gred{2.7} &  \gred{13.8}
\\
\multirow{2}{*}{{Normalization} ($1 \pm 0.05$)} 
& \gblue{10.5} & \gblue{1.8} & \gblue{3.1} & \gblue{13.1} & \gblue{3.5} & \gblue{0.9} & \gblue{0.9} & \gblue{3.6} \\[-1mm]
&  \gred{26.5} &  \gred{4.2} &  \gred{9.2} &  \gred{70.6} & \gred{11.1} &  \gred{2.1} &  \gred{3.7} & \gred{14.8}
\\
\hline
\multirow{2}{*}{Combined} 
& \gblue{ 9.3} & \gblue{1.7} & \gblue{2.8} & \gblue{12.4} & \gblue{3.4} & \gblue{0.8} & \gblue{0.8} & \gblue{3.5} \\[-1mm]
&  \gred{20.0} &  \gred{3.2} &  \gred{7.6} &  \gred{33.9} & \gred{ 8.6} &  \gred{1.9} &  \gred{1.9} & \gred{ 9.1}
\end{tabular}
\caption{The impact of systematic errors on the sensitivity to the octant of the 
				 atmospheric mixing angle for muon-like 
				 events only (blue) as well as both muon-like and cascade events (red) 
				 for different true values of the atmospheric mixing
				 angle and mass hierarchy, NH on the left and IH on the right, 
				 in 1-year run of PINGU.}
\label{tab:oct1}
\end{table}


For the octant determination, comparison between \gtab{tab:oct0} and 
\gtab{tab:oct1} shows also that the systematic errors 
in energy and angular resolutions, as well as that in the muon mis-identification
rate, do not significantly reduce the sensitivity. It drops significantly when 
the systematic error in the overall normalization is introduced, when only the 
muon-like events are considered. The sensitivity can be recovered again, once the
cascade events are included in the analysis.

\section{Conclusion}
\label{sec:conclusion}

The physics reach of measuring the atmospheric neutrino oscillation pattern
at PINGU to determine the neutrino mass hierarchy, the atmospheric mixing 
angle and its octant is explored in detail by making use of the decomposition 
property of the observable event rates in the propagation basis. Smearing in 
the reconstructed neutrino energy and the zenith angle due to the neutrino CC
scattering kinematics has been carefully studied, together with the energy and
angular resolutions of the detector. We find that the smearing effects reduce the
sensitivity to the neutrino mass hierarchy, by one order of magnitude while the 
precision on the atmospheric mixing angle and sensitivity to its octant is 
worsened by a factor around $2$. The mass hierarchy sensitivity can increase 
by up to $40\%$ if the muon-like events are split into two channels with the
criterion $1 - y \gtrless 0.67$ by estimating the inelasticity $y$ of each
event. It also improves slightly the precision on the atmospheric
mixing angle and its octant determination. These benefits from the inelasticity
measurement of the muon-like events can partially compensate the negative
effect of detector resolutions and muon mis-identification. Including the 
cascade events can not only increase the sensitivity of all the measurements, 
but most importantly stabilize the sensitivity against the influence from systematic 
errors in the muon mis-identification rate and the overall flux normalization.
Further improvements can be expected from detailed optimization of the muon
channel splitting criterion, or even refined binning of muon inelasticity
instead of splitting the muon events into just two channels, and including
the down-going events as well as the high energy atmospheric neutrinos to
reduce the flux uncertainties. We hope that our simple but systematic analysis 
will help preparing dedicated studies with full detector simulation.

\section{Acknowledgements}

The authors are grateful to the discussions with Francis Halzen and 
Carsten Rott on the PINGU 
detector and SFG would like to thank Costas Andreopoulos for kind help on 
GENIE. The Japan Society for the Promotion of Science (JSPS) has 
generously provided SFG a postdoc fellowship to do research at KEK,
which is deeply appreciated. 
This work is supported in part by Grant-in-Aid for Scientific research 
(No. 25400287) from JSPS.

\bibliographystyle{hunsrt}
\bibliography{pingu2}
\nocite{*}

\end{document}